# Ratio of Quantiles Indicates Burstiness with Fewer False Negatives than the Conventional Burstiness Parameter

Joshua Z. Stadlan[1,2,3,*], Michelle Birkett[1,2,4], Jason H. Rife[3]

[1] Center for Computational & Social Sciences in Health, Northwestern University, Chicago, IL, USA

[2] Northwestern Institute on Complex Systems, Northwestern University, Evanston, IL, USA

[3] Department of Mechanical Engineering, School of Engineering, Tufts University, Medford, MA, USA

[4] Department of Medical Social Sciences, Feinberg School of Medicine, Northwestern University, Chicago, IL, USA

[*] Corresponding author. Email: joshua.stadlan@kellogg.northwestern.edu

## Abstract

Complexity researchers view burstiness–fluctuating levels of activity–as evidence of hidden interactions within the system generating the activity signal. Yet, current burstiness metrics miss evidence of burstiness in some moderately bursty distributions and under moderate sampling conditions. The canonical Burstiness Parameter (BP) compares distributions of timing statistics to the exponential distribution, representing the timing of independent random events, but it provides false negatives for some parameter ranges of power laws, with and without cut-offs. We introduce a metric that maintains BP's measurement approach but reduces false negatives: the Burstiness Tail-based Index (BTI). Based on ratios of differences in quantiles, BTI correctly classifies bursty distributions over certain parameter ranges misclassified by BP. Additionally, we demonstrate BTI to be more robust than BP in the presence of limited sample sizes and short observation windows, using simulated samples drawn from distributions correctly classified by BP in their analytical form. As a case study, we revisit an analysis of human activity data and find that the choice of BTI over BP influences interpretations of the timescales of burstiness in the dataset. Given these analytical, simulated, and empirical results, we argue for BTI's practical advantage over BP in assessing burstiness in real-world temporal signals for complexity research and time series modeling.

## Introduction

In studying a system's activity pattern–whether an engineered, natural, or social system–a modeler wants to know to what extent individual events are inter-related. The modeler may ask: to re-create the pattern of gaps between events, is it necessary to understand the connections between events within the system, or can the associations within a system be ignored?

For example, an emergency call center may assume that at any given hour of a typical day, the pattern of incoming calls emerges from random independent emergencies of individuals [1], [2], [3], [4]. However, that assumption breaks down in a wider time range. In a wider time range, effects of inclement weather, major accidents, holidays, and large gatherings induce flurries of correlated activity, resulting in clustered emergency calls [5].

These "bursts" of call activity make emergency call rates on this broader scale an example of what researchers call "burstiness": the presence of irregularly fluctuating event frequencies or activity levels in a system [6]. Human interaction generally exhibits burstiness, as documented across a variety of scales and media – email exchanges, cell phone calls, face-to-face conversations, etc. [6], [7], [8].

Meanwhile, that typical hour of a typical day's emergency calls, discussed above, can be labeled "non-bursty" – they lack burstiness [9]. A pattern can also be "anti-bursty": consider the activity pattern of a supervisor calling the emergency center agents periodically for assessment purposes. Those calls are likely to be distributed more evenly than random. As such, the burstiness of a system's activity can be measured on a scale of anti-bursty to bursty, with non-bursty in the middle, using the distribution of its inter-event times. In this paper, we present a new metric of burstiness for empirical activity data, advancing current measures that compare a distribution of inter-event times to the distribution expected under the non-bursty case of unrelated, random individual events.

Since the burstiness of a time pattern may emerge from interactions of a system's components, it can serve as key initial evidence of complexity [10], [11]. A simple burstiness metric can help characterize signals from different domains, to identify large-scale trends in complex systems. Even for modelers concerned with forecasting or decision-making and not complex systems research, a burstiness metric can help them determine whether to approximate their pattern of interest with non-bursty assumptions (Poissonian process assumptions) to simplify their analysis. Due to the simplifying power of these approximations [12], engineers and data scientists traditionally start with these assumptions for queuing analysis [9] and reliability analysis unless they have compelling evidence for their subject patterns' deviation from it (e.g., [13], [14]). Researchers defend these assumptions in modeling the frequency of vehicle accidents at a fixed location [15] and debate using the assumptions in emergency dispatch staffing optimization analysis [4]. The risk of simplification is that an inappropriate non-burstiness assumption might skew conclusions about important real-world outcomes. The presence of burstiness in human contact patterns can, for example, determine whether a communicable disease outbreak dies out, explodes, or lingers endemically [16], [17], [18].

At present, there is no universally accepted metric for burstiness in the analysis of complex systems; sometimes burstiness is only defined phenomenologically [19]. We adopt the common approach that the burstiness of a system is best measured by the heaviness of the tail of the probability distribution of a key temporal output, such as inter-event time, compared to the exponential distribution, on an empirically relevant domain [20][21][22]. The exponential distribution is chosen as an apt reference point, as it describes the inter-event times of uncorrelated random events. Distributions with tails heavier than the exponential distribution are bursty. Distributions similar to the exponential distribution are non-bursty, and those with lighter tails are anti-bursty [22].

How to implement a readily quantifiable metric for real-world datasets is not obvious. Formal definitions of tail heaviness concern behavior in the limit to infinity, which is not directly observable in empirical samples. This limitation leaves empirical burstiness assessment open to multiple options of aspects of deviation from the exponential distribution to capture. When analyzing a single distribution of an inter-event time series, some researchers side-step the quantification question by simply visualizing the distribution and pointing out power law-like regions. This approach has replication challenges, due to many histogram visualization choices that influence the interpretation, and is limited to examining a few curves at a time [23][24]. In a comparative analysis of burstiness across many datasets (e.g. [19], [25], [26]), relying solely on visual analysis of histograms would be prohibitive. Moreover, in characterizing the burstiness of output from a computational model under different conditions or parameter spaces (e.g. [27], [28], [29],[30]), a quantitative metric is required to construct a response surface or estimate partial derivatives.

Of the limited efforts to directly quantify burstiness, the most referenced burstiness metric in the field of complexity is the Burstiness Parameter (henceforth, BP) of Goh and Barabási, which does adopt the exponential distribution-comparison approach [19]. Applications of BP include characterization of intensive-care-unit visit times [31], particle interactions [32], grocery checkout lines [33], metapopulation models [34], Wikipedia editing [22], and Python package creation [35]. However, as we will demonstrate, the BP metric does not always indicate heavy tails on certain parameter ranges for common distributions. Furthermore, BP converges slowly in terms of sample size, such that the metric can be unreliable, even when analyzing data sets of thousands of samples.

With the goal of more reliably characterizing temporal distributions' tail heaviness in signals extracted from complex systems, this paper introduces an alternative burstiness metric to BP: the Burstiness Tail-based Index (BTI). BTI is an adaptation of the Index of Tail Weight, a related metric from quality control engineering literature using quantile ratios [36]. We argue that BTI is more reliable than BP, both in detecting a wider range of common heavy tail distributions and in converging more rapidly for a limited quantity of data.

The rest of the paper is organized as follows. First, we define the BTI metric, with reference to the Index of Tail Weight, and describe the existing BP metric for comparison. Next, we compare BP and BTI on two fronts: accuracy on analytical distributions and robustness on sample data. On the analytical side, we present parameter ranges of a few key heavy tail distributions where BTI successfully agrees with our understanding of burstiness while BP does not. On the sample data considerations, we compare the metrics' convergence under moderate sample sizes and finite observation window limitations. We then present an example of a recent empirical burstiness analysis where the choice of burstiness metric influences the results. We conclude with a discussion of the strengths and weaknesses of the two metrics and their relationship to different facets of burstiness.

# Results

## Defining BTI and BP

In this section, we define our BTI metric by adapting the Index of Tail Weight (ITW). We then review the definition of BP.

The Index of Tail Weight (ITW), found in engineering statistics literature, compares the length of portions of a distribution's tail via three quantiles – at a reference probability $p_{ref}$, a close tail probability $p_{closeTail}$, and a farther tail probability $p_{farTail}$ [36]:

$$ITW_\Phi(F; p_{ct}, p_{ft}, p_{ref}) = \frac{\left(\frac{F^{-1}(p_{farTail}) - F^{-1}(p_{ref})}{F^{-1}(p_{closeTail}) - F^{-1}(p_{ref})}\right)}{\left(\frac{\Phi^{-1}(p_{farTail}) - \Phi^{-1}(p_{ref})}{\Phi^{-1}(p_{closeTail}) - \Phi^{-1}(p_{ref})}\right)} \quad \text{(eq. 1)}$$

where $F^{-1}$ is the inverse cumulative distribution function (cdf), or quantile function, $\Phi^{-1}$ is the inverse cdf of the standard normal distribution. By default, $p_{closeTail} = 0.75, p_{farTail} = 0.99, p_{ref} = 0.50$. This index, assuming a symmetric distribution, takes a measure of right tail length–here, ratio of the difference between 99[th] percentile and median, to the difference between 75th and median–and compares it to that of a standard normal distribution.

Figure 1 uses the Student's T Distribution for $\nu = 1$, which has heavier tails than the normal distribution, to illustrate these reference quantiles. Note that the ratio of the x-axis length of the 75th-99th percentile region (shaded) to the x-axis length of the 50th-75th percentile region (striped) is much larger for the Student's T distribution, in red, than for the standard normal distribution, in blue.

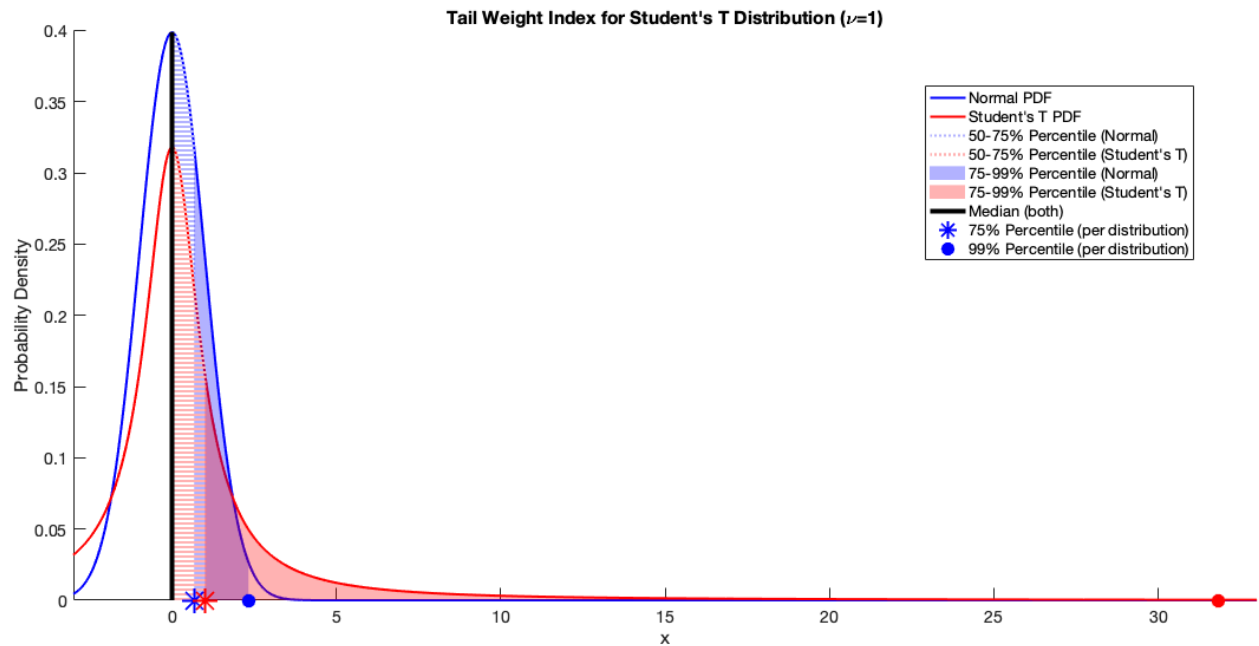

*Figure 1: Elements of the Tail Weight Index for Student's T Distribution. Student's T pdf is in red, the standard normal $\mathcal{N}(0,1)$ in blue. Striped area represents the 50-75th percentile for each distribution, and the solid shaded area represents the 75-99th percentile for each distribution. The 50th percentile (the median) is at x=0 for both, in the solid black line. Asterisk on the x-axis marks the 75th percentile and circle marks the 99th.*

ITW is defined in reference to a two-tailed distribution, the normal distribution; it assumes symmetry and only measures the right tail. Despite its inelegant definition, the ITW is straightforwardly interpretable: it answers, "how much longer is some key portion of the tail than normal?" The question for a burstiness measure is similar, except that the discriminant distribution is the exponential distribution instead of the normal distribution.

Thus, we can adapt $ITW_\Phi(F; p_{closeTail}, p_{farTail}, p_{ref})$ as a burstiness measure, replacing the comparison to the normal distribution $\Phi$ with a comparison to the exponential distribution $Exp$. When the tail ratio of a distribution equals that of an exponential distribution, $ITW_{Exp} = 1$. We map $ITW_{Exp}$ from $(0, \infty)$ to (-1,1), with 0 as the burstiness-discriminant[1]. The mapping corresponds to the burstiness scale for BP, as summarized in Table 1.

*Table 1: Interpretation of the Burstiness Scale*

| Burstiness Value | Description |
| --- | --- |
| $B \to -1^+$ | Maximally antibursty (periodic) |
| $-1 \leq B \ll 0$ | Antibursty: less dispersed than an exponential distribution |
| $B \approx 0$ | Non-bursty: equivalent to an exponential distribution |
| $0 \ll B \leq 1$ | Bursty: more dispersed than an exponential distribution |
| $B \to 1^-$ | Maximally bursty |

The resulting definition for BTI, based on ITW with an exponential distribution as the reference distribution, is as follows:

$$BTI(F; p_{closeTail}, p_{farTail}, p_{ref}) = \frac{ITW_{\text{Exp}}(F; p_{closeTail}, p_{farTail}, p_{ref}) - 1}{ITW_{\text{Exp}}(F; p_{closeTail}, p_{farTail}, p_{ref}) + 1} \quad \text{(eq. 2)}$$

We inherit the default parameters from ITW of $p_{closeTail} = 0.75$, $p_{farTail} = 0.99$, and $p_{ref} = 0.50$ for (eq. 2). Figure 2 illustrates the elements of BTI for two distributions: a light-tailed distribution, Half-Normal (top panel); and a heavy-tailed distribution, Lognormal (bottom panel). The striped and shaded regions help illustrate the length of the 50th-75th percentile and 75th-99th percentile regions, respectively, with the asterisk and circle shapes on the x-axis indicating the values that go into the BTI calculation.

---

[1] We adopt the $(0, \infty)$ to $(-1, 1)$ mapping for consistency with the Burstiness Parameter's mapping (defined below). However, we note that, as a ratio of *nested* lengths, $ITW_{\text{Exp}}$ has a positive infimum of $\inf ITW_{\text{Exp}} = \frac{\Phi^{-1}(p_{closeTail}) - \Phi^{-1}(p_{ref})}{\Phi^{-1}(p_{farTail}) - \Phi^{-1}(p_{ref})}$, such that the mapped parameter never approaches -1. For example, the extreme anti-bursty case of an overlapping 75th and 99th percentile (i.e., a probability point mass > 0.24 in the tail) would only result in a mapped value of approximately -0.7 (using the default $ITW_{\text{Exp}}$ parameters of $p_{ct} = 0.75$, $p_{ft} = 0.99$, and $p_{ref} = 0.50$). In this paper, we retained the nested-tail formulation for consistency with the established metric of ITW, but we note that using a ratio of non-overlapping lengths (e.g., $\frac{q_c - q_b}{q_b - q_a}$) would facilitate better use of the negative values for the metric. Alternatively, the ITW definition could be conserved and the mapping could be a piecewise function, where $ITW < 1$ is scaled to use the full range $(-1, 0)$.

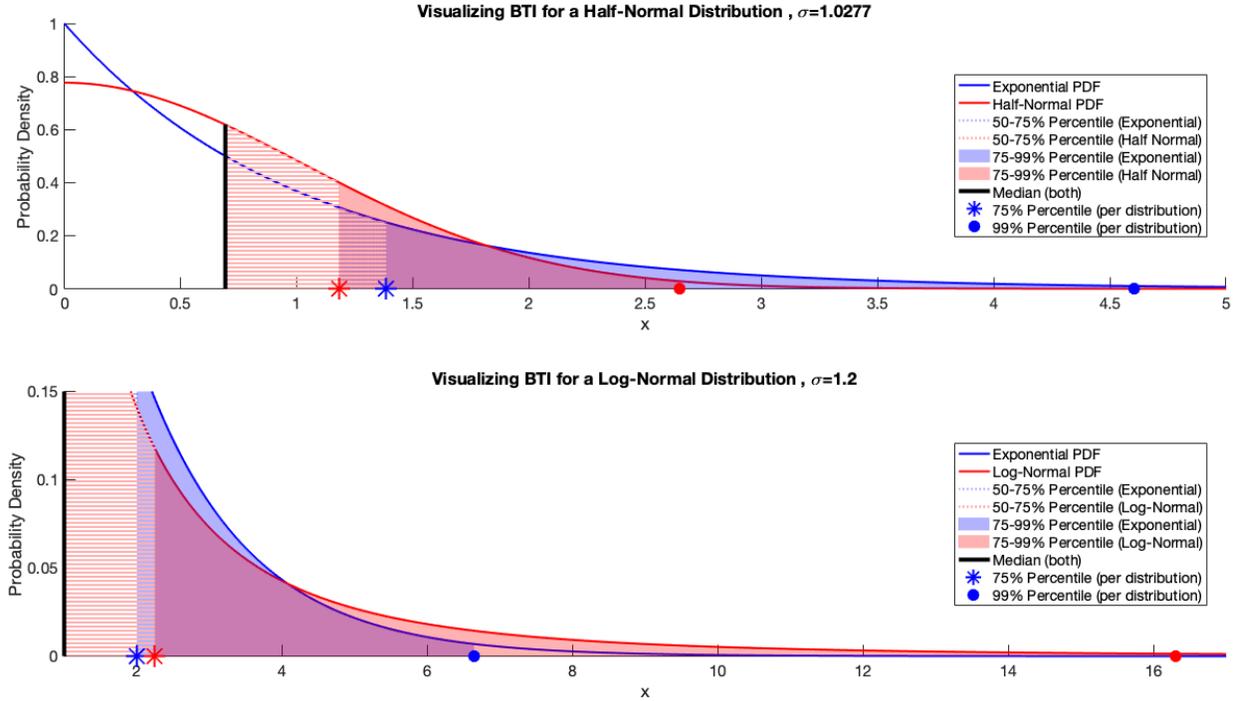

*Figure 2: Elements of the Burstiness Tail-based Index, for a canonical light-tailed (Half-Normal; top panel) distribution and heavy-tailed (Lognormal; bottom panel) distribution. The pdf under consideration is in red, the exponential distribution in blue. In both cases, distribution parameters were chosen such that the medians of the distribution under consideration and the reference exponential aligned. Striped area represents the 50-75th percentile for each distribution, and the solid shaded area the 75-99th percentile for each distribution. The 50th percentile (the median) is indicated by a solid black line. Asterisk on the x-axis marks the 75th percentile and circle marks the 99th.*

Unlike BTI, BP is based on the Coefficient of Variation, which is the ratio of standard deviation to mean: $CoV = \frac{\sigma}{\mu}$. BP maps $CoV$, which has a range of $(0, \infty)$, to the (-1,1) burstiness scale described above in the context of BTI:

$$BP = \frac{CoV - 1}{CoV + 1} = \frac{\sigma - \mu}{\sigma + \mu} \qquad \text{(eq. 3)}$$

The mapping implies that an infinite-variance time series is maximally bursty, with $BP = 1$, while a process with completely regular intervals–and therefore zero variance–is maximally anti-bursty, with $BP = -1$. Conveniently, $CoV = 1 \rightarrow BP = 0$ corresponds to the exponential distribution, since its mean and standard deviation are equal.

Of note, since a sample variance cannot be infinite, Kim and Jo propose a correction to allow for $BP = 1$ for a finite sample [21], which we apply in the small-sample section below.

## Analytical Comparison for Canonical Distributions

Several common distributions have parameter ranges for which BTI classifies burstiness appropriately and BP does not, without ranges where BP is correct and BTI is not. In this section we consider three such distributions commonly considered in bursty dynamics: the power law, the power law with exponential cut-off (PLEC), and the lognormal distribution. In this section, we use analytical computation on these canonical distributions to compute exact values for BP and BTI. Recall that we want a burstiness metric to return positive values to indicate burstiness, and therefore BP and BTI should always give positive values for these three burstiness-associated distributions.

The continuous power law distribution has probability $p(x) \propto x^{-\alpha}$ with $\alpha > 1$ and is the paradigmatic distribution for burstiness [22][37]. The power law begins at some minimal value $x_{min}$, resulting in a PDF of $f(x) = (\alpha - 1)x_{min}^{\alpha-1}x^{-\alpha}$, with support $x \in [x_{min}, \infty)$ [23][2]. For the power law distribution, BTI is unambiguously positive, whereas BP is not.

As shown in panel B of Figure 3, there are two regions of $\alpha$ that do not yield positive BP values. For $\alpha \leq 2$, the mean is undefined, so the analytical evaluation of BP is undefined. The other region of $\alpha$ yielding non-positive BP values is $\alpha \geq 2 + \sqrt{2}$. BP is 0 at $\alpha = 2 + \sqrt{2}$, as highlighted by the red dot in the figure, and decreases as $\alpha$ increases. While many power law-fitted human activity datasets estimate $\alpha$ values between 2 and 3 [23][6], several documented power laws have exponents in the negative-BP range. For example, Clauset et al. note sales of books (3.7), email address books size (3.5), and papers authored (4.3) [23]. Meanwhile, BTI yields positive values for all valid values of $\alpha$. It decreases monotonically with increase in $\alpha$, and its analytical limit converges to 0 as $\alpha \to \infty$, appropriate for an infinitely steep power law. Its limit is 1 as $\alpha \to 1^+$, appropriate for an infinitely heavy power law (We provide the derivation of these limits in Appendix A).

---

[2] Note we follow [23]'s parameterization for power law and power law with exponential cut-off, which prescribes the simpler parametric relationship for the pdf, as opposed to the Pareto distribution and tempered Pareto, which prescribes the simpler parametric relationship for the complementary cdf.

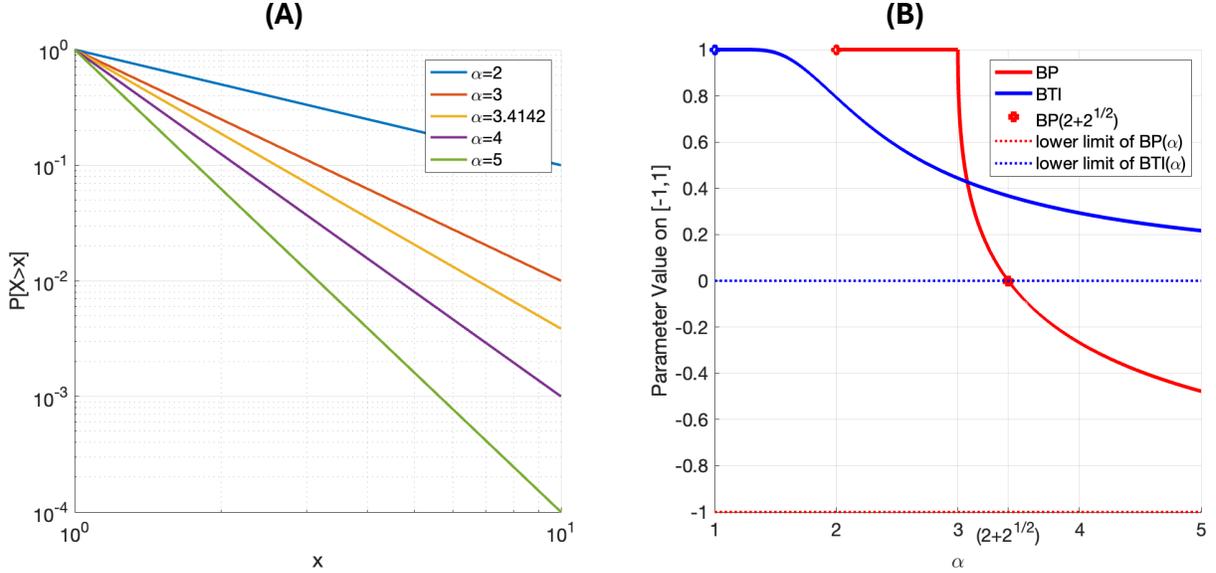

Figure 3: Comparing BP and BTI for a power law distribution with varying values of exponent α. Panel A: complementary cumulative density function (ccdf) for the power law distribution for varying exponent α, with $x_{min} = 1$. Panel B: BP (red) and BTI (blue) calculated over varying alpha (these values are independent of the value of $x_{min}$). BP is undefined below 2, equal to 1 for $2 < α ≤ 3$, and decreases below 0 for $α > 2 + \sqrt{2}$, approaching $-1$ as $α → ∞$. BP's intercept at 0 is indicated with a solid red circle. BTI approaches 1 as $α → 1^+$, and asymptotically approaches 0 as $α → ∞$, never crossing into negative. The lower limits of BP and BTI are indicated by red and blue dotted lines, respectively.

Similarly, for the PLEC distribution, BTI is again unambiguously positive, whereas BP is not. To appreciate this result, consider the form of the PLEC distribution, which is a generalized power law with a soft cut-off [38]:

$$p(x) = Cx^{-α}e^{-λx} \text{ for } α > 0, λ > 0, x \in [x_{min}, ∞) \quad \text{(eq. 4)}$$

where C is a normalization factor to ensure a valid probability distribution. We can nondimensionalize the random variable described by this distribution by scaling by $x_{min}$ such that the scaled random variable $Y$ always starts at $y = 1$:

$$Y = \frac{X}{x_{min}}, \; λ^* = λx_{min}, \; p(y) = C^*y^{-α}e^{-λ^*y} \text{ for } α > 0, y \in [1, ∞) \quad \text{(eq. 5)}$$

Cut-offs emerge in real-world processes from a constraint such as finite resources [39]; arguably constraints are nearly universal, so that power laws without cut-offs are at least rare [40] if not impossible [41]. For this practical reason, PLEC distributions are still generally treated as heavy-tailed (despite never exhibiting a strictly sub-exponential limit) and considered bursty [41],[42],[43].

Despite PLEC being associated with burstiness, BP values for PLEC decrease into negative values as $λ^*$ increases, misleadingly indicating anti-burstiness. By comparison,

BTI conforms to the expected behavior: BTI is positive for all valid PLEC distributions. We illustrate this in two ways. In Figure 4, we generate a heatmap of BP values for PLEC distributions over the parameter space $\alpha \in (0, 4] \times \lambda^* \in [10^{-4}, 10^0]$. Positive values of BP are colored in blue, and negative values in red, with lighter shades near zero. The black curve indicates where $BP_{PLEC}(\alpha, \lambda^*) = 0$. While many typical parameter combinations are captured in the positive (blue) region, some relevant regions, such as sub-regions of $\{2 < \alpha < 3, \lambda^* > 0.05\}$, yield negative BP. For example, Delabays and Tyloo find that the distribution of the number of articles by author in the journal *Science* between 1900-1940 is well-fitted to a discrete PLEC distribution with $\alpha = 2.13, \lambda^* = 0.09$ [44]. This is located just above the $BP = 0$ curve: $BP = -0.01$, misleadingly close to the BP value of an exponential distribution, while $BTI = 0.34$ appropriately identifies a bursty distribution.

To appreciate how BP and BTI respond differently to cut-off parameter changes, we plot BP and BTI values together in Figure 5, for varying $\lambda^*$, with fixed $\alpha = 2.1$. We selected this value of $\alpha$ since it is in the common range for naturally occurring datasets, and its corresponding unrestricted power law yields a maximally bursty BP value, 1.0. In panel (B), BTI is on the vertical axis and BP on the horizontal axis, while colors correspond to values of $\lambda^*$. Looking at the top right corner, note that they approach similar values as $\lambda^* \to^+ 0$: at $\lambda^* = 10^{-4}$, BP is 0.83 and BTI is 0.74. However, as $\lambda^*$ grows and exponential behavior begins to dominate the shape of the distribution, the metrics diverge: BTI approaches 0 (non-bursty), but BP approaches -1 (anti-bursty). For reference, recall that a purely exponential distribution always has a burstiness of BP=0, BTI=0. Note that the BP and BTI relationship is not a line but an S-curve: BTI is more descriptive than BP for the $\lambda^*$ range on the order of $10^{-2} - 10^{-1}$, a common parameter range for PLEC use (see, e.g., the parameter fits in [44]). BP, on the other hand, is more sensitive to $\lambda^*$ changes than BTI is for $\lambda^* < 10^{-3}$ and $\lambda^* > 1$, outside the more common parameter ranges to describe empirical datasets.

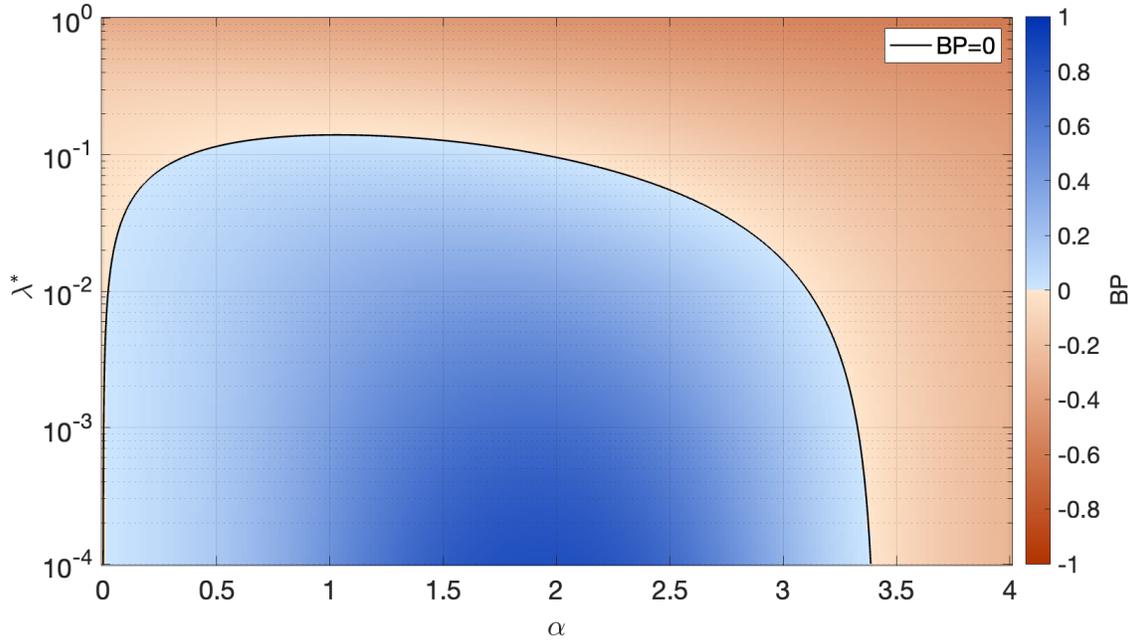

Figure 4: BP heatmap for PLEC distributions as a function of their parameters $\lambda^*$ (on the vertical axis) and $\alpha$ (on the horizontal axis). Colors correspond to the value of $BP(PLEC(\alpha, \lambda^*))$, with their lightness/darkness corresponding to absolute value $|BP|$. Positive values are colored in blue, and negative values in red, with lighter shades near zero. The black curve indicates where $BP(PLEC(\alpha, \lambda^*)) = 0$.

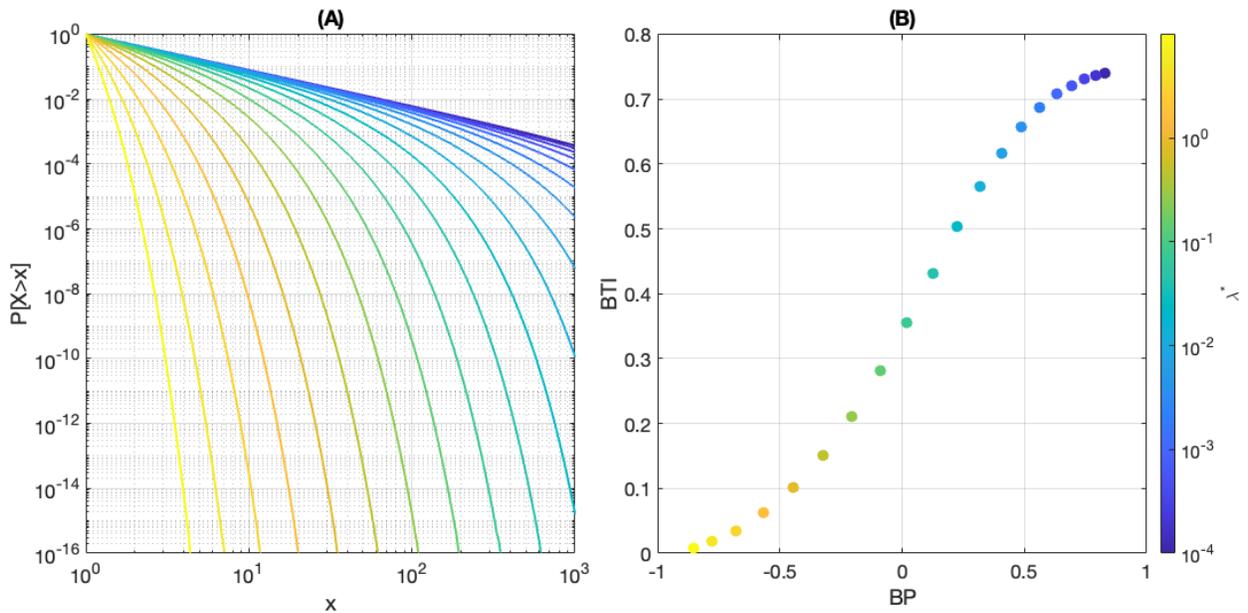

Figure 5: Burstiness of the PLEC distribution. (A) CCDF of the PLEC distribution, at $\alpha = 2.1$, colored by value of $\lambda^*$ (see color bar on the right side of the figure). (B) Values of BTI (vertical axis) and BP (horizontal axis) for the PLEC distribution, at $\alpha = 2.1$ and colored varying $\lambda^*$ corresponding to the CCDF curves on the datasleft panel.

The lognormal distribution is significant as another bursty distribution common to temporal datasets [45], [46], [47], [48], [49]. For example, it can model the time it takes for an exponential-growth process to reach a threshold size, where the threshold varies centrally (i.e., similar to a normal distribution) [49]. The following equation describes the lognormal distribution:

$$P(x) = \frac{1}{x\sigma\sqrt{2\pi}} \exp\left(-\frac{(\ln x - \mu)^2}{2\sigma^2}\right) \qquad \text{(eq. 6)}$$

For the lognormal distribution, neither BTI nor BP is unambiguously positive; however, BTI remains positive for a wider parameter range. The key distribution parameter for the coefficient of variation is $\sigma$: $CoV = \sqrt{\exp(\sigma^2) - 1}$ [50]. Consequently, as recognized previously by Goh and Barabási [19], BP yields negative values for any $\sigma < \sqrt{\ln(2)} \sim 0.83$. It decreases steeply as $\sigma$ decreases below this point, with BP($\sigma = 0.1$) = -0.82, close to periodic (-1). BTI decreases less steeply than BP, as illustrated by the BTI vs. BP curve in the right panel of Figure 6. By comparison, BTI takes negative values only below a lower crossover point (for $\sigma < \sim 0.53$).[3] While the domain where BTI correctly indicates heavy tail and BP does not is limited, $\sim 0.53 < \sigma < \sim 0.83$, there are real-world distributions that occur in this range. For example, Barcelo and Jordan fit message holding times to a lognormal distribution with $\sigma = 0.7$ [51], and Hurley et al. fit Vela pulsar microglitch waiting times to $\sigma = 0.79$ [52], [53].

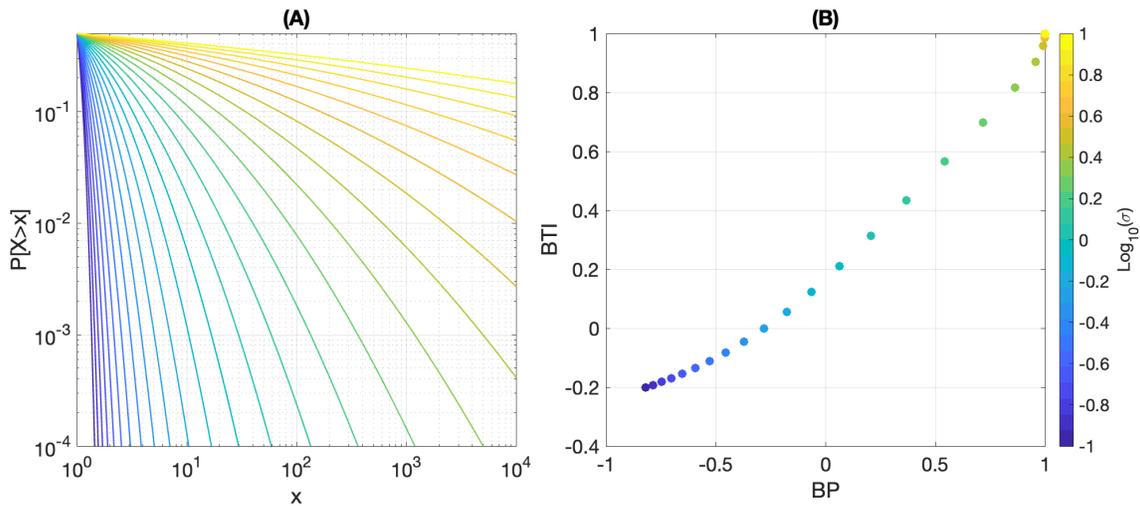

Figure 6: Burstiness of the lognormal distribution. (A) CCDF of the lognormal distribution at $\mu = 0$, colored by value of $\sigma$ (see color bar on the right side of the figure). (B) Values of BTI (vertical axis) and BP (horizontal axis) for the lognormal distribution, at $\mu = 0$ and colored by varying $\sigma$, corresponding to the CCDF curves in (A).

---

[3] Choosing reference quantiles farther in the tail for BTI reduces this cut-off slightly, but at the expense of reasonable quantiles for empirical datasets.

## Sample-Based Convergence Comparison

In sequential statistical computation of BP and BTI, using Monte Carlo simulations as described in the Methods section, we observe that BTI converged to the correct result much more quickly than BP. We consider two effects: sample size effects and windowing effects.

Small sample sizes are well known to yield poor estimates of a metric concerned with tail behavior, especially for distributions with thick tails, since events many orders of magnitude longer than the mean time occur with non-negligible probability, wildly affecting the outcome [21]. Still, it is desirable that a burstiness metric converges quickly to the true analytical result when analyzing real-world data collections, where the sample size is bounded, for instance, through collection of timing quantities (e.g. event times and inter-event times) within some finite window of time $T_f$.

Windowing effects skew the sample distribution from the true underlying generating distribution toward shorter events (an effect which can, to some extent, be modeled, as discussed in Appendix B). However, with a sufficiently large number of non-windowed samples, the basic BP and BTI metrics should converge to their theoretical infinite-time values, as described above for the power-law, PLEC, and lognormal distributions.

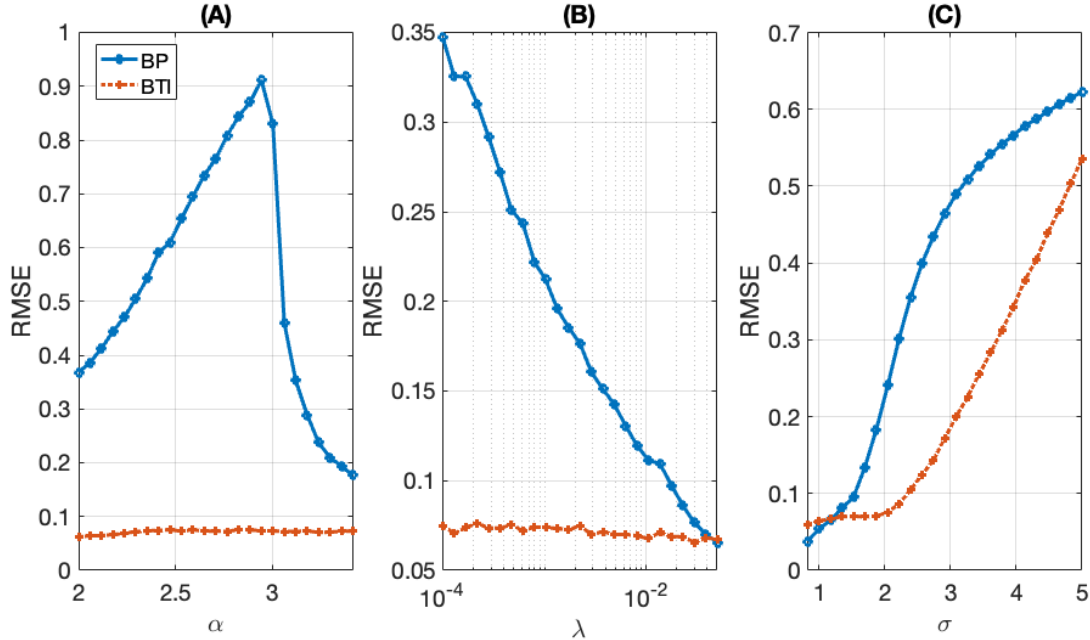

*Figure 7: Root-Mean-Squared Error (RMSE) for BP and BTI on simulated data from three distributions at typified sample size $N = 10^3$, within simulated window $T_f = 10^3$, over $M = 10^3$ runs, relative to their analytical values from the uncensored analytical distributions. (A) Power law distribution with $x_{min} = 1$ ; (B) PLEC with $x_{min} = 1$, $\alpha = 2.5$; (C) Lognormal with $\mu = 1$. Parameter ranges shown are the subsets of the canonical ranges presented in [24] for which the sign of BP is positive.*

In Figure 7, we plot root mean squared error (RMSE) of sample BP and BTI calculated on $10^3$ samples representing observations of event durations within a $T_f = 10^3$ time window, over $10^3$ Monte Carlo runs. These simulated events begin at start times distributed uniformly within the window, and their "true" durations may get censored by the end of the window. (Left-censoring–where events may be in progress at the start of the observation window–is beyond the scope of this paper.) We generate the "true" event durations from (A) a power law, (B) PLEC, and (C) lognormal distributions, and sweep over a distribution parameter for the subsets of the canonical ranges presented in [24] for which the sign of BP is positive. For most of the relevant parameter ranges (where BP and BTI agree on burstiness as a binary indicator) in this windowed scenario, BP exhibits larger errors than BTI. In other words, sample BTI more robustly recovers the underlying uncensored analytical distribution's BTI value.

To isolate the effects of sample size and observation window, we fix the underlying distribution's parameters in Figure 8 and separately sweep over sample size (left column) and observation window (right column). In (A)-(C), the mean of sample BTI hugs its true value for sample sizes above $10^2$, while the mean of sample BP remains negatively biased past $10^3$. While sample BTI has larger 75% confidence interval for truly small sample sizes (under $10^2$), the uncertainty region reduces to smaller than that of sample BP by $10^3$ in all three cases. In (D)-(F), considering window sizes, BTI converges to its true value at much smaller window sizes relative to BP in all three cases as well.

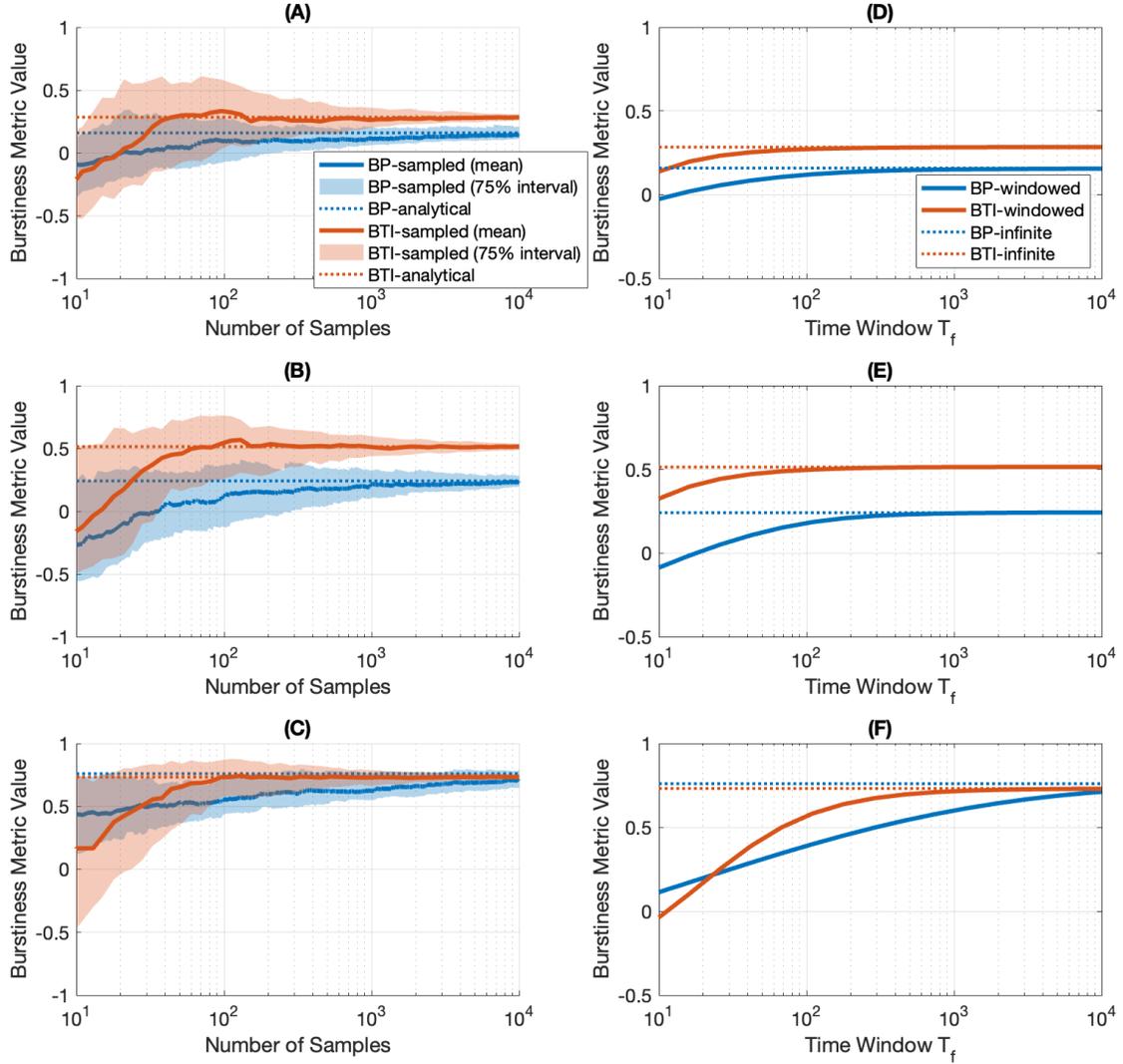

*Figure 8: Left panels A-C: Convergence comparison of sample BP (blue) and sample BTI (red) relative to their true values (dotted lines) with increasing sample sizes, mean in solid line and shaded 75% uncertainty region by quantiles; Right panels D-F: Convergence comparison of sample BP (blue) and sample BTI (red) relative to their true values (dotted lines) with increasing time window size. The time window unit is in multiples of the corresponding distribution's median (e.g., 10x the median, 100x the median, etc.). Distribution parameters for all panels: (A & D) Power Law: $\alpha = 2.1$, (B & E) PLEC: $\alpha = 2.5, \lambda^* = 10^{-2}$, (C &F) Lognormal: $\mu = 1, \sigma = 2$.*

## Empirical Case Study

In this section, we revisit a recent empirical burstiness analysis and apply the BTI metric. We searched for a recent peer-reviewed article with "burstiness" in the contribution, where BP is utilized, and where the temporal data analyzed are made available. We identified Takeuchi and Sano's work measuring the burstiness of human movement in everyday activities by presenting positive BP values of inter-event times in their data [54]. They collect activity data from wearable accelerometers on children while playing and adults while resting, performing housework, and performing desk-based work. They find, via BP, stable burstiness for activity defined as changes in scalar acceleration (jerk magnitude) above ~ 100 ms$^{-3}$, supported by appropriate fits to power laws with exponential cut-off. However, they observe that the inter-event times for lower thresholds are "close to the exponential distribution of a stationary Poisson process" because of random noise (see Takeuchi and Sano's Appendix B).

We calculate BTI on Takeuchi and Sano's datasets for varying jerk magnitude threshold $\theta$. Figure 9 shows the burstiness level of Inter-Event Times (IETs) for jerk at various θ for four activities: child play, deskwork, housework, and rest time. The figure reveals a divergence between BP and BTI: we find decisively positive BTI values for even the lowest of jerk magnitude thresholds registered by the sensors, in contrast to the Poissonian-like indication (near zero) for low thresholds.

Specifically, in panel (B) of the figure ($\theta > 30 \ m/s^3$), BTI largely agrees with BP, with a small positive shift in the burstiness scale. This can be seen in the linear trend between BTI and BP in the right panel of Figure 9. Here, higher $\theta$ corresponds to both higher BTI and BP. BTI scales roughly linearly with BP, with a positive offset.

However, the figure reveals that the metrics diverge in their burstiness assessments in panel (A), for $\theta < 30 \ m/s^3$. Here, BTI continues to detect burstiness across all settings, while BP only does so for the child play dataset. Deskwork has a consistently bursty value in BTI; rest is extremely bursty in BTI for $\theta < 9 \ m/s^3$, and mildly bursty between $10 - 20 \ m/s^3$, but anti-bursty in BP for this entire range. Housework is mildly bursty in BTI but mildly anti-bursty in BP. Only child play is bursty according to both BP and BTI.

The corresponding ECCDFs (empirical complementary cdf) provide more information to help determine whether the data are actually bursty or not. In Figure 10, we plot the ECCDFs for each activity against a reference curve: the best-fit exponential distribution (dotted lines). The data clearly diverge from the exponential distribution in all cases (for all thresholds of jerk), which is more consistent with the BTI indication of burstiness than the near-zero values of BP. (Model fit statistics corroborate the ill-fit of the exponential distribution, and PLEC parameter estimates on this data predict the misleading BP indication – see Appendix D.) The fluctuating, negative BP values for this jerk magnitude threshold range conceal the robustly bursty signature of human movement at these scales, as visible in Figure 10.

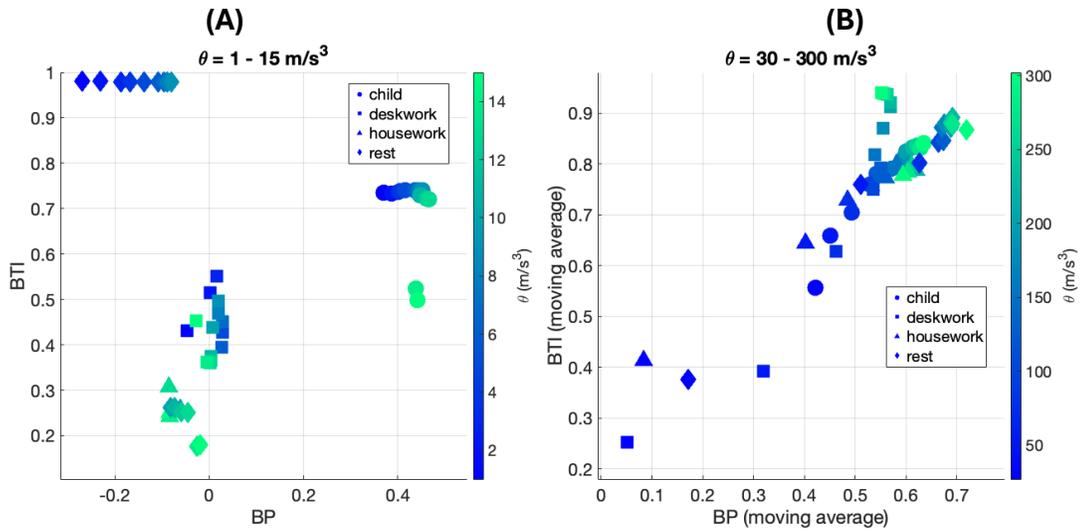

Figure 9: *IETs defined at varying θ thresholds on four human activity datasets from [54], where color indicates the threshold value and symbol indicates the dataset. (A) Lowest range of $\theta$: $1 - 15\ m/s^3$, at increments of 1. (B) Broad range $\theta = 30 - 300\ m/s^3$ at increments of 20. Note the general agreement (linear trend) between BTI and BP in (B), but extreme disagreement in (A): some values are highly bursty in BTI (close to 1) but anti-bursty (below 0) in BP. Either BP or BTI is providing a misleading indication of burstiness level for the activity signals at low filter thresholds.*

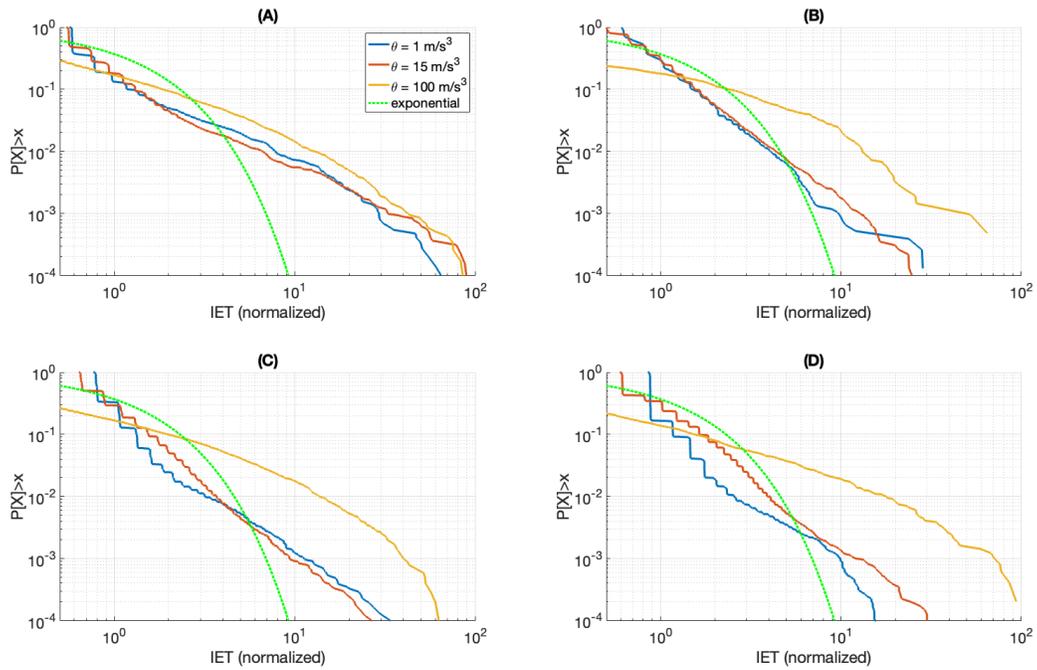

Figure 10: *Empirical complementary CDFs for human activity IETs at various jerk magnitude thresholds θ, distinguished by color, for the four activity datasets: (A) child play, (B) deskwork, (C) housework, (D) rest. The x axis represents multiples of the mean IET of each dataset (i.e., the datasets were normalized by their respective means).*

# Discussion

We have shown several advantages to BTI over BP in indicating the exponential-comparison notion of burstiness in empirical signals, with zero as the discriminant. First, we demonstrated that BTI conforms with our desired behavior for power laws and PLEC, while BP does not for large exponents ($\alpha > 3.4$) as well as large $\lambda$ for PLEC (threshold provided in Figure 4). BTI also correctly classifies a greater parameter range for lognormal distribution than BP, though it still exhibits a gap. On simulated sample data, we surveyed performance under typical sample sizes and observation windows, finding that sample BTI generally achieved greater convergence to its analytical value than BP.

For many empirical datasets, BP and BTI will agree on the level of burstiness. However, there are important cases where they do not. We presented a scenario from a recently published study where the choice of metric does in fact impact the data analysis, and checking the full distributions visually supported the BTI-based indication of burstiness over the BP indication of non-burstiness. However, always checking visually is not an option for an objective comparison across many datasets or sensitivity analysis of a computational model. For those contexts that require a primary quantitative metric for batch comparison or sensitivity study, BTI is better suited than BP.

## Comparison to BP Alternatives

While we focused our comparison on BTI to BP, we speculate that the advantages of BTI highlighted here will hold against several other candidate burstiness metrics as well. In the Introduction we justified our comparison to BP based on both its wide adoption and simplicity. Simplicity is a good reason for adoption: data analysts regularly calculate the mean and variance when introduced to a new dataset; mean-normalized variance (coefficient of variation) is a natural extension. We argue that BTI, in comparing quantiles, is similarly advantageous in only calculating elemental quantities. A lead competing approach to quantifying burstiness, using bursty "train" sizes, requires adoption of the bursty train concept, the design choice of train window, and then still requires a statistical summary metric of the distribution of train size [10]. That choice of summary metric could still be a quantile-based ratio like BTI.

Another alternative approach to BTI and BP, anticipated by Goh and Barabási in their paper defining BP, is using a measure directly from the pdf or CDF of the distribution [19]. We prefer a quantile (inverse CDF)-based ratio due to the numerical stability and greater information range of working with ratios of large numbers (quantiles in the tails) compared to pdf values (less stability) or cdf values (narrower information range) in the tails, especially when working with empirical data. Indeed, stability of quantile ratios is a theme in our simulated data results. It is not surprising that BTI is more reliable than BP for measuring burstiness on samples from canonical distributions, as estimating sample quantiles is more reliable than estimating moments for heavy-tailed distributions. Adding one extreme value outlier to a moderate number of samples may not change BTI very much, if samples already exist close to the right of the existing 75% and 99% quantiles, but it will likely substantially increase BP, given the effect of $\frac{1}{N}x^2$ on the variance. Therefore, we encourage the use of BTI instead of BP for potentially heavy-tailed sample data, which is where a burstiness test is most often of interest.

## Limitations

There are several drawbacks to BTI. While not a concern for typical probability distributions, BTI is sensitive to small changes in distribution shape that occur near its reference quantiles. This can be a problem for distributions with one or more modes in the tails, such as timing data of events that mostly occur unscheduled and organically but are occasionally scheduled at long, fixed intervals. We explore this sensitivity in Appendix C.

Even within the confines of the common burstiness quantification approach adopted here–of comparing the tail of the probability distribution of a key temporal output to the exponential distribution–there are still multiple possible ways in which the tail could diverge from the exponential distribution, and no self-evident burstiness "ranking" of these divergences. For several examples: (a), the tail could start with a power law-like region but follow an exponential curve beyond this region. Or, vice versa, (b), a curve could start exponential but transitions to sub-exponential in the limit of $x \to \infty$, fulfilling the strict definition of "long tail" [55]. Or, (c), an irregular distribution (e.g. piecewise and/or multi-modal) that clearly diverges from an exponential distribution in terms of pdf curve fit, yet the value of its cdf at modes or discontinuities might still fit an exponential distribution. In other words, one could ask about at least three senses of divergence corresponding to these examples:

(a) is there a power law-like region in the tails?

(b) Is the limit $x \to \infty$ sub-exponential?

(c) is there much more probability mass in the tails than expected under an exponential distribution?

Often these senses align, but not always. One such corner case where these senses do not align is the PLEC distribution. PLEC fulfills (a) and (c) but not (b). We have been most concerned with (a) and secondarily (c) in this paper, and least concerned with (b). We argue that (b) is least applicable to empirical data from finite observation windows, as it is always possible that the tail $x > \hat{x}_{max}$ changes, and evaluating the limit requires parametric model fit assumptions.

Sometimes (a) and (c) compete as well: looking back to our case study in Figure 10, we observed that, as we decreased the threshold $\theta$ thereby shortening the IETs, the power law-behavior becomes more pronounced–becoming burstier by (a)–even as the probability mass overall shifts left, becoming less bursty by (c) (though still bursty). For a parametric distribution example: comparing a power law with a very high $\alpha$ or a stretched exponential distribution with very low $\beta$, which looks like $x^{-1}$ for the first several decades.

We have shown that BTI is well-suited at addressing (a) and secondarily (c). BTI picks up on some distribution shape (i.e. decay rate) comparisons by approximating the slope of unit of $x$ over probability mass, providing evidence for (a). For common parametric distributions, comparing the decay rate at the, for instance, 75% and 99% percentiles, is a good indicator of the decay rate at higher $x$ and avoids the irrelevant information of the probability mass at the head of the distribution. Additionally, focusing on quantiles means the metric remains tied to general location of probability mass (certainly for functions that are non-decreasing past the median, like the common parametric distributions), addressing (c). Therefore, the quantile-ratio approach encoded in BTI responds to both the distribution shape (a) and relative probability mass location (c) for well-behaved distributions, but it is not informative about (b).

As a moment-based measure, BP is more sensitive to the behavior at $x \to \infty$. However, as shown in the power law analysis, it misclassifies even parametric power laws at high $\alpha$. Researchers sympathetic to the mathematical definition of heavy tails, working on parametrically modeling extremely large datasets with extremely large domains, can bypass both metrics and just assess the limit analytically based on their parametric model fits. But in the more common empirical dataset context, BTI better captures the more intuitive, empirical sense of burstiness.  (See Appendix E for a proposed limit-based definition of burstiness that better aligns with (a) and (c).)

In our discussion of power laws and PLEC, we demonstrated that BTI responds as expected to change in the parameter values of bursty distributions. As the curve steepens (higher exponent $\alpha$) or the exponential cut-off shifts leftward (higher rate $\lambda$), BTI appropriately decreases, while remaining positive. However, we did not investigate BTI and BP's performance in rankings of relative burstiness between different bursty parametric distributions. Returning to the conceptual example above, we might want to compare a power law with a very high $\alpha$ or a stretched exponential distribution with very low $\beta$, which looks like $x^{-1}$ for the first several decades. Given the different senses of burstiness, such a ranking would require further definition, which we leave for future work. One potential research direction is to assess the predictive power of burstiness metrics on the relative magnitude of a known burstiness-related effect, since, as mentioned in the introduction, the effects of burstiness are some of the motivators of a burstiness metric in the first place. For example, relative burstiness levels among bursty distributions could be pegged to the reduction of the epidemic threshold on a contact network [16], [17], [18]. Such a validation approach–of evaluating burstiness metrics by its predictive performance on a burstiness-influenced outcome–could allow for a pragmatic comparison of burstiness metrics that differ in their quantitative definition of burstiness (such as the bursty train size approach [10]), which is beyond the scope of this paper.

      We have introduced a simple quantile-based metric for burstiness, BTI. We have demonstrated its accuracy, stability, and interpretability in rapid detection of burstiness in sample data. BTI is more accurate than the conventional metric, BP, in indicating burstiness when temporal data follows a power law-like behavior, which is precisely when burstiness is most meaningful to complexity science researchers. BTI should be especially preferred over BP in the following scenarios: (1) when a power law with exponential cut-off is one of the candidate distributions, or (2) in empirical sample scenarios when the sample mean and variance are unstable and (3) the data is significantly censored by an observation time window. In our case studies, we find that the conditions of sample BP inaccuracy (relative to BTI) may occur if the number of samples is between $10^2 - 10^4$ or the time window is less than ~$10^4$ times larger than the median event duration. The quantification of burstiness could benefit from further work in consolidating a consensus definition, in validating metrics via prediction accuracy, and hypothesis testing to interpret sample BTI values near zero (e.g., using bootstrapping). In the meantime: BTI is a simple, robust, easily interpretable indicator of burstiness for unimodal empirical and simulation temporal data to help identify temporal evidence of complexity.

# Method

## Sample-based Convergence Comparison

To compare the convergence of BP and BTI on sample data to their analytical values, we generated synthetic data from power law, PLEC, and lognormal distributions. PLEC and lognormal are continuous distributions presented in Clauset et al.'s power law tutorial as potential alternative hypotheses in assessing whether empirical data fits a power law [23], as well as in Barabási's textbook as common heavy-tailed distributions in network science [56].

The synthetic data was generated using MATLAB function random and Clauset et al.'s randht function for sampling from PLEC [23], simulated as time-observations within an observation window $T_f$, with $N$ samples per observation set. To simulate windowed data for $T_f$, where an event has uniform probability of starting at any time $0 \leq t < T_f$, we generated unwindowed samples and took the minimum of a simulated "remaining time" in the window, approximating $X_{windowed}$ as $min(X_{unwindowed}, T_f - T_{start})$ where $T_{start} \sim Uniform(0, T_f)$. We simulated $M = 10^3$ windows per distribution in the evaluation set. We selected $N = 10^3$ samples and $T_f = 10^3$ as a reasonable lower bound to the dataset characteristics of the typical face-to-face interaction temporal network dataset from the SocioPatterns Collaboration, which has on the order of $10^4$ contact events and span a window of ~$10^4$ time units [57]. We calculated BTI and BP on each $M$ simulated observation set and evaluated mean-squared error relative to their analytical value. We picked 25 parameter values for each dataset, spanning the range of examples presented in [24] for which the sign of BP is appropriately positive (i.e., we excluded from consideration heavy-tail ranges where BP is incorrectly negative). For PLEC, we swept over $\lambda$ since the trend of $\alpha$ is already explored for the power law. For lognormal, we swept over $\sigma$ since that is the only parameter that BP and BTI depend on (analytically).

To generate Figure 8, we separated the sample size effect from the windowing effect. On the left side, we generated M=100 datasets of unwindowed, $N = 10^4$ samples, and plotted BP and BTI on subsets of the M datasets of increasing n from 10 to $10^4$. The right panels evaluate BP and BTI on $N = 10^6$ samples transformed by $x'_i = \min(x_i, T_f v : v \sim U(0,1))$ at 25 log-spaced values of η in $T_f = \eta Med(X)$, where $Med(X)$ is the median value of the unwindowed distribution.

## Empirical Case Study

For the empirical case study, we use the data described in [54] and apply their processing rules as follows. We first determine the signal by thresholding the magnitude of the jerk vector:

$$E(t) = \begin{cases} 1 & J(t) \geq \theta \\ 0 & J(t) < \theta \end{cases} \text{ where } J(t) = \frac{1}{\Delta t}\sqrt{\Delta a_x(t)^2 + \Delta a_y(t)^2 + \Delta a_z(t)^2} \qquad \text{(eq. 7)}$$

We then calculate the IETs from recorded times, $t_{event}$, where $E(t_{event}) = 1$. To match Takeuchi and Sano's procedure, we (1) ignore gaps larger than 60 seconds, (2) ignore segments with more than 30 gaps of 10 seconds or longer, and (3) exclude IETs under 0.3 seconds; we then combine the resulting IETs of the five recordings for each of the four settings (child play, deskwork, housework, and rest).

# References

[1] R. Larson and A. Odoni, "4.6 Center for Emergency Calls: Queueing Systems of the Birth-and-Death Type," in *Urban Operations Research*, Online Edition., Hoboken, NJ: Prentice-Hall, 1981, ch. 4.

[2] N. Gans, G. Koole, and A. Mandelbaum, "Telephone Call Centers: Tutorial, Review, and Research Prospects," 2003, *INFORMS Inst.for Operations Res.and the Management Sciences*. doi: 10.1287/msom.5.2.79.16071.

[3] Ľ. Jánošíková, P. Jankovič, M. Kvet, and F. Zajacová, "Coverage versus response time objectives in ambulance location," *Int. J. Health Geogr.*, vol. 20, no. 1, Dec. 2021, doi: 10.1186/s12942-021-00285-x.

[4] D. S. Matteson, M. W. McLean, D. B. Woodard, and S. G. Henderson, "Forecasting emergency medical service call arrival rates," *Annals of Applied Statistics*, vol. 5, no. 2 B, pp. 1379–1406, Jun. 2011, doi: 10.1214/10-AOAS442.

[5] P. L'Ecuyer, K. Gustavsson, and L. Olsson, "Modeling Bursts in the Arrival Process to an Emergency Call Center," in *Proceedings of the 2018 Winter Simulation Conference*, Gothenburg, Sweden: IEEE, Dec. 2018, pp. 525–536. Accessed: Nov. 25, 2025. [Online]. Available: doi:10.1109/WSC.2018.8632536

[6] M. Karsai, H.-H. Jo, and K. Kaski, "Bursty Human Dynamics," Mar. 2018, doi: 10.1007/978-3-319-68540-3.

[7] P. Holme and J. Saramäki, "Temporal networks," *Phys. Rep.*, vol. 519, no. 3, pp. 97–125, 2012, doi: 10.1016/j.physrep.2012.03.001.

[8] P. Holme and J. Saramäki, "Temporal Network Theory," *Computational Social Sciences*, pp. 1–24, 2019, doi: 10.1007/978-3-030-23495-9_1.

[9] B. C. Moss, "On the Distribution of Inter-Arrival Times of 911 Emergency ResponseProcess Events," 2020. Accessed: Mar. 11, 2026. [Online]. Available: https://scholarsarchive.byu.edu/etd/8391

[10] M. Karsai, K. Kaski, A. L. Barabási, and J. Kertész, "Universal features of correlated bursty behaviour," *Sci. Rep.*, vol. 2, 2012, doi: 10.1038/srep00397.

[11] D. Marković and C. Gros, "Power laws and self-organized criticality in theory and nature," Mar. 10, 2014. doi: 10.1016/j.physrep.2013.11.002.

[12] L. Green, "Queueing Analysis in Healthcare," in *Patient Flow: Reducing Delay in Healthcare Delivery. International Series in Operations Research & Management Science*, vol. 91, R. Hall, Ed., Boston, MA: Springer, 2006, ch. 11, pp. 281–307. Accessed: Mar. 11, 2026. [Online]. Available: https://doi.org/10.1007/978-0-387-33636-7_10

[13] A. Heinen, "Modelling Time Series Count Data: An Autoregressive Conditional Poisson Model," *SSRN Electronic Journal*, 2011, doi: 10.2139/ssrn.1117187.

[14] P. J. Diggle, "Spatio-temporal point processes: methods and applications," in *Statistical Methods for Spatio-Temporal Systems, Monographs on Statistics and Applied Probability*, no. June 2005, 2006.

[15] A. Nicholson and Y.-D. Wong, "Are Accidents Poisson Distributed? A Statistical Test," *Accidents Analysis & Prevention*, vol. 25, no. I, pp. 9–97, 1993, doi: https://doi.org/10.1016/0001-4575(93)90100-B.

[16] M. Akbarpour and M. O. Jackson, "Diffusion in networks and the virtue of burstiness," *Proc. Natl. Acad. Sci. U. S. A.*, vol. 115, no. 30, pp. E6996–E7004, Jul. 2018, doi: 10.1073/pnas.1722089115.

[17] T. Takaguchi, N. Masuda, and P. Holme, "Bursty Communication Patterns Facilitate Spreading in a Threshold-Based Epidemic Dynamics," *PLoS One*, vol. 8, no. 7, Jul. 2013, doi: 10.1371/journal.pone.0068629.

[18] M. Mancastroppa, A. Vezzani, M. A. Muñoz, and R. Burioni, "Burstiness in activity-driven networks and the epidemic threshold," *Journal of Statistical Mechanics: Theory and Experiment*, vol. 2019, no. 5, May 2019, doi: 10.1088/1742-5468/ab16c4.

[19] K. I. Goh and A. L. Barabási, "Burstiness and memory in complex systems," *EPL*, vol. 81, no. 4, Feb. 2008, doi: 10.1209/0295-5075/81/48002.

[20] T. Hiraoka, N. Masuda, A. Li, and H. H. Jo, "Modeling temporal networks with bursty activity patterns of nodes and links," *Phys. Rev. Res.*, vol. 2, no. 2, Apr. 2020, doi: 10.1103/PhysRevResearch.2.023073.

[21] E.-K. Kim and H.-H. Jo, "Measuring burstiness for finite event sequences," Apr. 2016, doi: 10.1103/PhysRevE.94.032311.

[22] M. Karsai and H.-H. Jo, "Measuring and Modeling Bursty Human Phenomena," Dec. 2024, [Online]. Available: http://arxiv.org/abs/2412.13617

[23] A. Clauset, C. R. Shalizi, and M. E. J. Newman, "Power-law distributions in empirical data," *SIAM Review*, vol. 51, no. 4, 2009, doi: 10.1137/070710111.

[24] Albert-László Barabási, *Network Science*, 1st ed. Cambridge, UK: Cambridge University Press, 2016.

[25] E. G. Altmann, J. B. Pierrehumbert, and A. E. Motter, "Beyond word frequency: Bursts, lulls, and scaling in the temporal distributions of words," *PLoS One*, vol. 4, no. 11, Nov. 2009, doi: 10.1371/journal.pone.0007678.

[26] H. H. Jo, T. Hiraoka, and M. Kivelä, "Burst-tree decomposition of time series reveals the structure of temporal correlations," *Sci. Rep.*, vol. 10, no. 1, Dec. 2020, doi: 10.1038/s41598-020-68157-1.

[27] J. Stadlan, *Comparing Explanations for the Burstiness of Face-to-Face Social Interaction via Homogeneous Agent-Based Models of Temporal Networks*, 31770415th ed., vol. 2025. Medford, MA: Tufts University ProQuest Dissertations & Theses, 2025.

[28] W. Zhong, Y. Deng, and D. Xiong, "Burstiness and information spreading in the active particles systems," Oct. 2022, [Online]. Available: http://arxiv.org/abs/2210.09139

[29] G. García-Pérez, M. Boguñá, and M. Á. Serrano, "Regulation of burstiness by network-driven activation," *Sci. Rep.*, vol. 5, May 2015, doi: 10.1038/srep09714.

[30] S. Unicomb, G. Iñiguez, J. P. Gleeson, and M. Karsai, "Dynamics of cascades on burstiness-controlled temporal networks," *Nat. Commun.*, vol. 12, no. 1, Dec. 2021, doi: 10.1038/s41467-020-20398-4.

[31] L. Bonomi and X. Jiang, "A mortality study for ICU patients using bursty medical events," in *Proceedings - International Conference on Data Engineering*, IEEE Computer Society, May 2017, pp. 1533–1540. doi: 10.1109/ICDE.2017.224.

[32] W. Zhong, Y. Deng, and D. Xiong, "Burstiness and information spreading in active particle systems," *Soft Matter*, vol. 19, no. 16, pp. 2962–2969, Mar. 2023, doi: 10.1039/d2sm01470j.

[33] A. Ahrens, O. Purvinis, D. Hartleb, J. Zaščerinska, and D. Micevičiene, "Analysis of a business environment using burstiness parameter: The case of a grocery shop," in *PECCS 2019 - Proceedings of the 9th International Conference on Pervasive and Embedded Computing and Communication Systems*, SciTePress, 2019, pp. 49–56. doi: 10.5220/0007977600490056.

[34] E. Fonseca Dos Reis and N. Masuda, "Metapopulation models imply non-Poissonian statistics of interevent times," *Phys. Rev. Res.*, vol. 4, no. 1, 2022, doi: 10.1103/PhysRevResearch.4.013050.

[35] A. Angdembe *et al.*, "bursty_dynamics: A Python Package for Exploring the Temporal Properties of Longitudinal Data," Nov. 2024, [Online]. Available: http://arxiv.org/abs/2411.03210

[36] D. Hoaglin, F. Mosteller, and J. Tukey, *Understanding Robust and Exploratory Data Analysis*. New York and Chichester: John Wiley and Sons, 1982.

[37] A. L. Barabási, "The origin of bursts and heavy tails in human dynamics," *Nature*, vol. 435, no. 7039, 2005, doi: 10.1038/nature03459.

[38] R. Pastor-Satorras and A. Vespignani, "Epidemic dynamics in finite size scale-free networks," *Phys. Rev. E Stat. Phys. Plasmas Fluids Relat. Interdiscip. Topics*, vol. 65, no. 3, 2002, doi: 10.1103/PhysRevE.65.035108.

[39] I. Voitalov, P. Van Der Hoorn, R. Van Der Hofstad, and D. Krioukov, "Scale-Free Networks Well Done," *Physics Review Research*, vol. 1, no. 3, p. 033034, Oct. 2019, doi: 10.1103/PhysRevResearch.1.033034.

[40] A. D. Broido and A. Clauset, "Scale-free networks are rare," *Nat. Commun.*, vol. 10, no. 1, Dec. 2019, doi: 10.1038/s41467-019-08746-5.

[41] C. Shalizi, "Chaos, Complexity, and Inference Lecture 13: Heavy-Tailed Distributions, Especially Power Laws," Feb. 2008. [Online]. Available: http://www.santafe.edu/~aaronc/powerlaws/

[42] T. Maschberger and P. Kroupa, "Estimators for the exponent and upper limit, and goodness-of-fit tests for (truncated) power-law distributions," *Mon. Not. R. Astron. Soc.*, vol. 395, no. 2, pp. 931–942, 2009, doi: 10.1111/j.1365-2966.2009.14577.x.

[43] J. Alstott, E. Bullmore, and D. Plenz, "Powerlaw: A python package for analysis of heavy-tailed distributions," *PLoS One*, vol. 9, no. 1, Jan. 2014, doi: 10.1371/journal.pone.0085777.

[44] R. Delabays and M. Tyloo, "Heavy-tailed distribution of the number of papers within scientific journals," *Quantitative Science Studies*, vol. 3, no. 3, pp. 776–792, Jun. 2022, doi: 10.1162/qss_a_00201.

[45] R. J. Lawrence, "The Lognormal as Event-Time Distribution," in *Lognormal Distributions*, 2018. doi: 10.1201/9780203748664-8.

[46] J. Gaddum, "Lognormal Distributions," *Nature*, no. 156, pp. 463–466, Oct. 1945, doi: 10.1038/156463a0.

[47] G. Auricchio, M. Ghiotto, S. Gualandi, and G. Toscani, "Generalized Galton's Boards Explain Social Phenomena via Statistical Physics."

[48] N. Blenn and P. Van Mieghem, "Are human interactivity times lognormal?," Jul. 2016, [Online]. Available: http://arxiv.org/abs/1607.02952

[49] E. Limpert, W. A. Stahel, and M. Abbt, "Log-normal distributions across the sciences: keys and clues," *Bioscience*, vol. 51, no. 5, pp. 341–352, 2001, [Online]. Available: http://stat.ethz.ch/vis/log-normal

[50] J. H. Proost, "Calculation of the Coefficient of Variation of Log-Normally Distributed Parameter Values," Sep. 01, 2019, *Springer International Publishing*. doi: 10.1007/s40262-019-00760-6.

[51] J. Jordan and F. Barcelo, "Statistical modelling of transmission holding time in PAMR systems," in *Conference Record / IEEE Global Telecommunications Conference*, 1997. doi: 10.1109/glocom.1997.632524.

[52] K. J. Hurley, B. McBreen, M. Rabbette, and S. Steel, "The lognormal properties of the soft gamma-ray repeater SGR 1806-20 and the Vela pulsar," *Astron. Astrophys.*, vol. 288L, pp. 49–52, 1994.

[53] J. I. Katz, "Log-Normal Waiting Time Widths Characterize Dynamics," *The Open Journal of Astrophysics*, no. 7, Jun. 2024, doi: 10.33232/001c.118582.

[54] M. Takeuchi and Y. Sano, "Burstiness of human physical activities and their characterisation," *J. Comput. Soc. Sci.*, vol. 7, no. 1, pp. 625–641, Apr. 2024, doi: 10.1007/s42001-024-00247-w.

[55] S. Foss, D. Korshunov, and S. Zachary, *An Introduction to Heavy-Tailed and Subexponential Distributions*, 2nd ed. New York, NY: Springer, 2013. doi: 10.1007/978-1-4614-7101-1.

[56] Albert-László Barabási, *Network Science*, 1st ed. Cambridge, UK: Cambridge University Press, 2016.

[57] M. Génois and A. Barrat, "Can co-location be used as a proxy for face-to-face contacts?," *EPJ Data Sci.*, vol. 7, no. 1, p. 11, 2018, doi: 10.1140/epjds/s13688-018-0140-1.

[58] M. Starnini, A. Baronchelli, and R. Pastor-Satorras, "Modeling human dynamics of face-to-face interaction networks," *Phys. Rev. Lett.*, vol. 110, no. 16, 2013, doi: 10.1103/PhysRevLett.110.168701.

[59] C. S. Gillespie, "Fitting Heavy Tailed Distributions: The poweRlaw Package," *J. Stat. Softw.*, vol. 64, no. 2, pp. 1–16, Mar. 2015, doi: 10.18637/jss.v064.i02.

## Acknowledgments

This article draws on work from the first author's doctoral dissertation, *Comparing Explanations for the Burstiness of Face-to-Face Social Interaction via Homogeneous Agent-Based Models of Temporal Networks* [27]. We thank Dr. Pratap Misra, Dr. Lenore Cowen, and Dr. Matt Koehler for serving on his dissertation committee and for helpful feedback on the earlier stage of this research.

## Author Contributions

J.S. and J.R. conceived of the study. J.S. developed the code and conducted the computational analyses. J.R. supervised. J.S. and J.R. wrote the main text. M.B. contributed to the Introduction and Discussion. All authors interpreted results, edited the narrative, and reviewed the final manuscript.

## Competing Interests

The authors declare no competing interests.

## Funding

No external funding was received for this research.

## Data Availability

All code required to reproduce the figures and analyses in the manuscript results will be available at https://github.com/jstadlan-compass/burstiness-tail-index and will be deposited in Zenodo upon publication. This study did not involve the analysis of new data sets. Please contact the corresponding author for any questions about running the MATLAB code.

# Appendix A

**Derivation of BTI Limits**

In this section, we derive the limits of BTI for power law distributions as $\alpha$ approaches infinity and approaches 1 from the right.

## I. BTI Limit of Power Law Distributions as $\alpha \to \infty$:

Recall that $BTI = \frac{r-1}{r+1}$, where $r = \frac{n}{d} = \dfrac{\dfrac{F^{-1}(p_{farTail})-F^{-1}(p_{ref})}{F^{-1}(p_{closeTail})-F^{-1}(p_{ref})}}{\dfrac{F_{exp}^{-1}(p_{farTail})-F_{exp}^{-1}(p_{ref})}{F_{exp}^{-1}(p_{closeTail})-F_{exp}^{-1}(p_{ref})}}$

At $p_{farTail} = 0.99$, $p_{closeTail} = 0.75$, and $p_{ref} = 0.5$, the denominator of $r$ is, given $F_{exponential}^{-1}(x) = -\frac{\ln(1-p)}{\lambda}$:

$$d = \frac{-\ln(1-0.99) - \ln(2)}{-\ln(1-0.75) - \ln(2)} = \log_2 50$$

Given the quantile function for power law distributions:

$$F_{powerLaw}^{-1}(x) = x_{min}(1-p)^{-\frac{1}{\alpha-1}}$$

and letting $t = \frac{1}{\alpha-1}$,

numerator $n$ becomes

$$n = \frac{100^t - 2^t}{4^t - 2^t}$$

In the variable of $t$, the limit $\alpha \to \infty$ becomes $t \to 0^+$:

$$\lim_{t \to 0^+} n = \left.\frac{100^t \ln(100) - 2^t \ln(2)}{4^t \ln(4) - 2^t \ln(2)}\right|_{t=0} = \log_2 50$$

Therefore,

$$\lim_{\alpha \to \infty} BTI_{powerLaw} = \frac{\frac{\log_2 50}{\log_2 50} - 1}{\frac{\log_2 50}{\log_2 50} + 1} = 0$$

Given $n(t) > \log_2 50$ for all finite $t > 0$, we conclude that BTI is always positive for all valid finite values of $\alpha$.

## II. BTI Limit for Power Law Distributions as $\alpha \to 1^+$:

As before, we have $BTI = \frac{r-1}{r+1}$, where $r = \frac{n}{d}$, and $n = \frac{F^{-1}(p_{farTail}) - F^{-1}(p_{ref})}{F^{-1}(p_{closeTail}) - F^{-1}(p_{ref})}$

Simplifying $n$,

$$n = \frac{100^t - 2^t}{4^t - 2^t} = \frac{50^t - 1}{2^t - 1}$$

As $\alpha \to 1^+$, $t \to \infty$ so we take the limit of $n$ as $t \to \infty$:

$$\lim_{t \to \infty} \frac{50^t - 1}{2^t - 1} = \infty$$

Since $\lim_{r \to \infty} \frac{r-1}{r+1} = 1$, therefore $\lim_{\alpha \to 1^+} BTI_{powerLaw} = 1$

# Appendix B

**Observation Window Effects**

(This appendix is excerpted from the first author's dissertation [27].)

An observation window has two censoring effects on samples of timing quantities, like event durations and inter-event times:

1. Durations cannot be longer than the time window $(\tau \leq T_f)$.
2. Many events, gaps, etc. that are ongoing at the end of the simulation / time window would have persisted for longer had the simulation / time window been extended. Similarly, if the beginning of the time window did not coincide with the start of any events, many events, gaps, etc. that are present at the start of the time window may have started earlier, had the simulation / time window been extended earlier in time.

Logically, these window effects disproportionately reduce the number of longer durations in the sample, skewing the sample distribution from the true underlying generating distribution toward shorter events.

We consider two approaches to working with temporal samples within finite observation windows: window-matching and probabilistic imputation.

## Window-matching

If the goal is for a simulation's output to match empirical datasets, without making claims about the underlying probability distribution if unconstrained in time, then the simplest approach is to simply set the simulation time to correspond to the empirical dataset's duration. This is the approach implicitly adopted in simulation-based studies such as [58]. The logic is, if the simulation would generate a matching distribution to the real-world distribution in the context of unconstrained time, then the window-censored simulated data will also match the window-censored empirical data. Whatever effect window-censoring has on the simulated sample would also be present in the empirical dataset.

For this correspondence to hold, how the time censoring is performed must be consistent across the empirical and simulated datasets. Namely, the censoring choice is between **early event termination** and **persisting event exclusion**. In early event termination, the analysis considers the end of the time window to be the end time of any ongoing durations. For example, if an event starts at $t = 88$ and the model window ends at $t = 100$, then the event duration corresponding to that event is from 88 to 100 even if the event were to persist if the simulation were extended in time. Similarly, if an event ends at $t = 75$, and the simulation ends at $t = 100$, then its inter-event time would be from 75 to 100 even if no new events were to start if the simulation were extended in time. In persisting event exclusion, events that persist at the conclusion of the simulation are disregarded. In the example above, the event that started at $t = 88$ is not recorded in the sample, nor is the inter-event time that began at $t = 75$.

It is worthwhile to note that while unconstrained-time correspondence implies window-censored correspondence, the converse is not strictly true. Equivalence of data under a known time window censoring does not imply that the data-generating functions would be equivalent in a time-unconstrained condition, or a different time window. Considering the timing of the durations and/or experimenting with different time windows makes spurious equivalences of this nature less likely.

## Probabilistic Imputation

The window-matching approach leaves unanswered, how does one estimate the uncensored distribution? An analyst can "unfurl" the uncensored distribution by parametrically modeling the time window censoring (under some assumptions).

If events of any duration are equally likely to start at any given time, then events of duration $x$ within the time window $[0, T)$ can only start in $\frac{T-x}{T}$ fraction of the window, yielding a probability mass function:

$$P_{censored}(x) = c \frac{T-x}{T} P_{uncensored}(x); \; 0 \leq x < T$$

Where $c$ is the normalizing constant to ensure $P_{censored}(x)$ is a proper probability distribution. The uncensored distribution can be uncovered under these assumptions via $P_{uncensored}(x) = \frac{T}{c(T-x)} P_{censored(x)}$ for $x < T$. Consider a censored version of the Pareto distribution (continuous Power Law with minimum $x_m$), with $x_m = 1$:

$$P_{censored\,Pareto}(x) = c \left(\frac{T-x}{T}\right) \alpha x^{-\alpha-1}$$

Integrating the censored pdf, and ensuring $F(T) = 1$, yields:

$$F_{censored\,Pareto}(x) = \frac{T(1-\alpha)}{T(1-\alpha) + \alpha - T^{1-\alpha}} \left[(1 - x^{-\alpha}) - \frac{\alpha}{T(1-\alpha)}(x^{1-\alpha} - 1)\right]$$

The censored and uncensored complementary CDF for a Pareto distribution with $x_{min} = 1, \alpha = 0.5$, are plotted in Figure 11. Note how the tail looks visually similar to exponential decay despite being the sum of two power terms. In recovering the uncensored CDF from the censored CDF, however, no values can be obtained beyond $T_f$ ($T_f = 10^3$ in the figure) without additional assumptions.

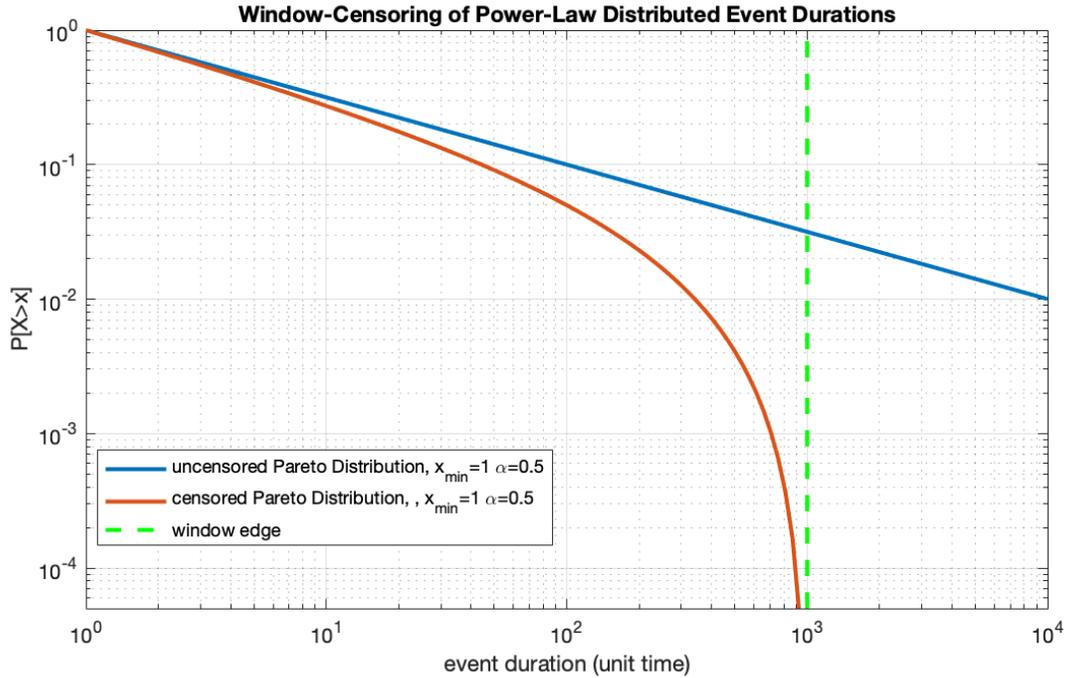

Figure 11: Theoretical effect of a time window on the observed samples from a Pareto distribution (power law)

This "uncensoring" method can be useful for correcting empirical data. However, if a researcher is looking to compare an infinite-time model to censored empirical data, best practices in data science would motivate applying a censoring to the model, and then comparing the censored version to the empirical data, to avoid hiding model assumptions in the data preparation. This implements a parametric version of the "window-matching" approach above.

# Appendix C

**BTI Sensitivity to Multi-Modal Distributions**

BTI is sensitive to the behavior of a distribution around its reference quantiles. This sensitivity affects the robustness of BTI metrics for distributions with a mode in the tail. To illustrate this sensitivity, we consider a mixture of two lognormal distributions:

$$X \sim w \, \text{lognormal}(\mu_1, \sigma_1^2) + (1-w) \, \text{lognormal}(\mu_2, \sigma_2^2) \qquad 0 \leq w \leq 1$$

Recall that lognormal distributions are always heavy-tailed, but that BP misclassifies lognormal distributions with $\sigma < 0.83$ as anti-bursty and BTI similarly misclassifies those under $\sigma < 0.53$. We select $\sigma_1 = 1$ for the first component of the mixture, corresponding to lognormal distributions that both metrics identify as bursty. We select $\sigma_2 = 0.1$ for the second component of the mixture, corresponding to lognormal distributions that both metrics identify as anti-bursty. We also fix $\mu_1 = 0$.

By varying $\mu_2$, we can adjust the location of the mode contributed by the second component, as seen in the pdf plots of five such parameter sets in panel A of Figure 12. By varying the relative weighting of the two components through $w$, we can adjust the where the mixture's inflection point (integral of the pdf mode) occurs on the quantile axis: in panel B, we plot the CCDF of the five distributions, and denote the 0.5, 0.75, and 0.99 quantiles in different broken line patterns.

Note that the inflection point of set 1 (blue), set 2 (red), and set 4 (purple) occur between the 0.75 and 0.99 quantiles; the inflection point of set 5 (green) occurs in between the 0.5 and 0.75 quantiles; and that of set 3 (yellow) occurs earlier (higher on y axis in CCDF) than the 0.5 quantile (median). Despite sharing the same parametric form, these distributions result in very different BTI values, as seen in Table 2. Sets 4 (purple) and 5 (green) have similarly located modes (2 vs. 3) and both are heavily weighed to the first component, and therefore share a similar BP value. Yet, because their inflection points are on opposite sides of the 0.75 quantile–the "close tail" reference quantiles for BTI–their BTI values are 0.63 apart.

We illustrate this BTI sensitivity, relative to BP, over a broader range of $\mu_2$ and $w$ values in the heatmaps in Figure 13. Both BP and BTI change signs over this parameter range. Yet, BP (panel A) for the most part exhibits gradual changes and mostly monotonic changes moving away from the $BP(\mu_2, w) = 0$ curve. BTI (panel B), on the other hand, exhibits rapid changes along the $BTI(\mu_2, w) = 0$ curve, moving from highly bursty to highly antibursty in a difference of about 0.5 in $\mu_2$. The BTI plot are more distinct changes in burstiness direction, as well as pools of constant burstiness value with sharp edges. For example, see the extremely bursty region on the bottom left quadrant and the anti-bursty region in the top right quadrant. The stark changes are due to inflection points close to the BTI reference quantiles; otherwise, BTI is more stable.

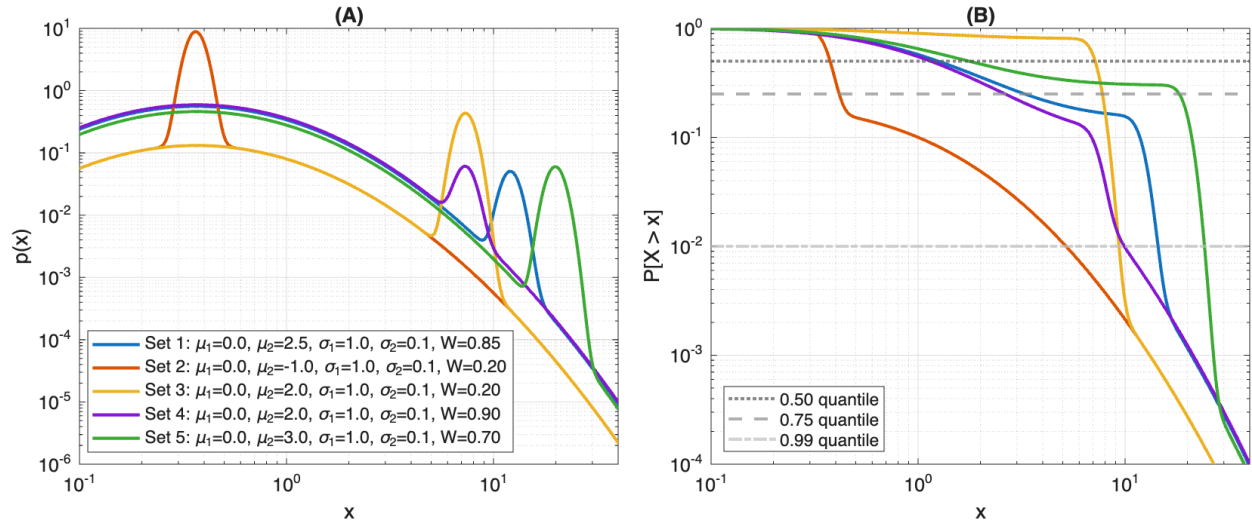

*Figure 12: Five parameter sets of two-component lognormal mixture distributions, plotted in log-log scale in solid lines and differentiated by color. (A) pdf curves for each parameter set. (B) CCDF curves for each parameter set, with the 50th, 75th, and 99th percentiles (reference quantiles for BTI) indicated by the grey broken lines.*

*Table 2: BP and BTI values for five parameter sets of two-component lognormal mixtures with the following fixed parameters: $\mu_1 = 0, \sigma_1 = 1, \sigma_2 = 0.1$*

| Set | $\mu_2$ | w | BP | BTI |
|---|---|---|---|---|
| 1 | 2.5 | 0.85 | 0.14 | 0.07 |
| 2 | -1.0 | 0.20 | 0.27 | 0.91 |
| 3 | 2.0 | 0.20 | -0.42 | -0.22 |
| 4 | 2.0 | 0.90 | 0.10 | 0.02 |
| 5 | 3.0 | 0.70 | 0.10 | -0.61 |

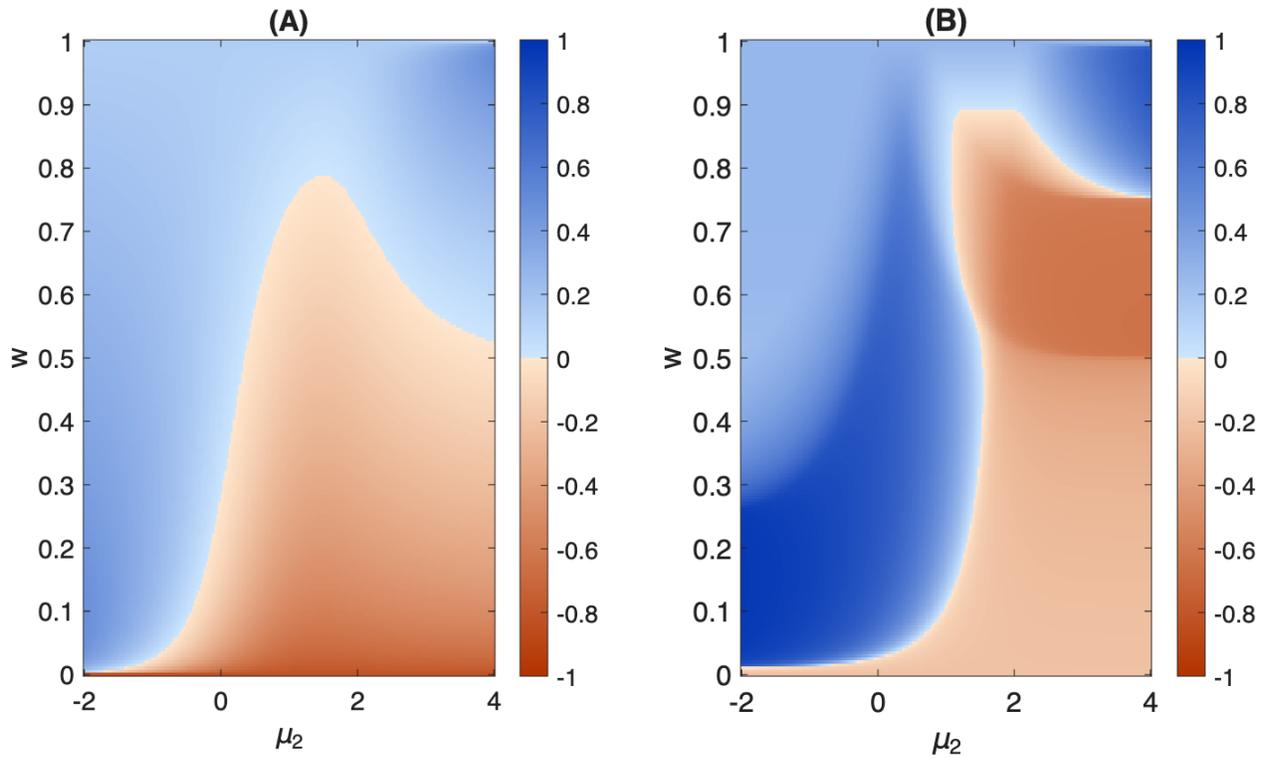

Figure 13: Heatmap of BP (A) vs. BTI (B) for a two-component lognormal mixture, with varying weight w and parameter $\mu_2$; the remaining parameters are fixed at $\mu_1 = 0, \sigma_1 = 1, \sigma_2 = 0.1$. Colors correspond to the value of the burstiness metric, with their luminosity corresponding to their absolute value. Positive values are colored in blue, and negative values in red, with lighter shades near zero.

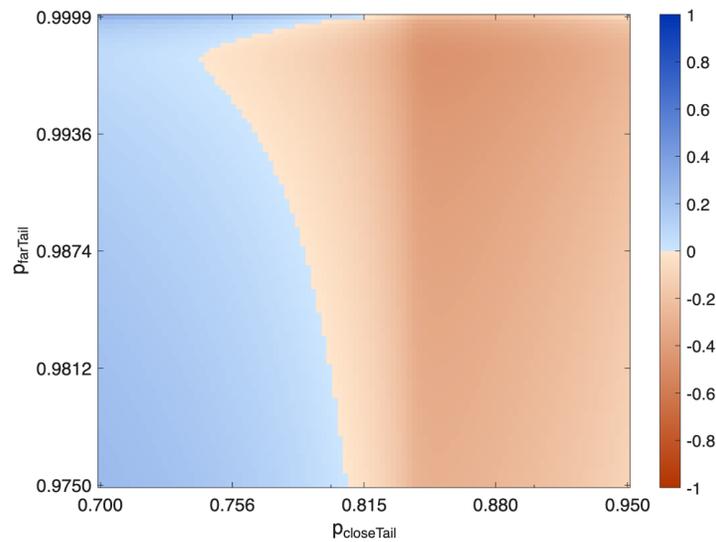

Figure 14: Heatmap of BTI for a two-component lognormal mixture distribution (parameter Set 1, from above) over varying choices of the BTI reference quantiles, $p_{closeTail}$ (x axis) and $p_{farTail}$ (y axis). Positive BTI values are colored in blue, and negative values in red, with luminosity corresponding to their absolute value (lighter shades near zero).

Another way to explore BTI's sensitivity around its reference quantiles is to keep the distribution fixed, and change the reference quantiles. We present this view in Figure 14: for Set 1 ($\mu_2 = 2.5, w = 0.85$), we display a heatmap of BTI values as a function of choice of $p_{closeTail}$ (x axis) and $p_{farTail}$ (y axis), i.e. $BTI(p_{closeTail}, p_{farTail})$. Within our example domain $[0.7, 0.95] \times [0.975, 0.9999]$, BTI ranges from -0.46 at $BTI(0.85, 0.9976)$ to 0.28 at $BTI(0.7, 0.9999)$, all describing the same distribution (Set 1). Granted, this is a contrived pathological case: designed so that moving the reference quantile $p_{closeTail}$ from 0.75 to 0.9 changes, as seen in Figure 12 panel B, whether the CCDF's inflection point (mode) is above or below the reference quantile. An investigation into how often these pathological cases occur naturally among distributions and datasets of interest in bursty dynamics is beyond the scope of this work.

# Appendix D

**Statistical Fits to Takeuchi & Sano Data**

To perform IET model fit to heavy-tail distributions, we follow Clauset et al.'s recommended "recipe" of (1) performing a continuous power law distribution fit on the data itself, (2) assessing, via comparing the observed KL-divergence with the power law fit-based simulated one, if the power law fit can be rejected with $p = 0.1$, then (3a) performing pairwise normalized likelihood ratio tests among reasonable alternative distribution models (with p values reported from Vuoung's test), with the same estimated $x_{min}$ from the corresponding continuous power law fit [23]. We add a step (3b): redoing the pairwise fit comparisons from (3a), but with a fixed $x_{min} = 0.3$ to force consideration of all of the datapoints (recall that this threshold was already applied in Takeuchi and Sano's data cleaning step, so there are no data points below 0.3 [54]).

We adapt Clauset and Shalizi's MATLAB and R powerlaw code, Alstott et al.'s Python powerlaw package [43], and Gillespie's poweRlaw package [59] to perform steps 1-3b. We process the activity datasets at $\theta = 1 \ m/s^3$, a low threshold where one might expect significant noise.

While a power law hypothesis was not consistent with the child play, housework, housework, and rest datasets, it could not be eliminated for deskwork ($p_{PL} = 0.31$). Yet, BP is just below zero (-0.05) for this dataset. The BTI value of 0.43 better reflects this power law plausibility result.

All four datasets at $= 1 \ m/s^3$ have higher likelihoods of power law and PLEC fits compared to their exponential fits, for power law estimated $x_{min}$, as reported in Table 3.

*Table 3: Normalized Log-Likelihood Ratios of Power Law and PLEC fits to the corresponding exponential distribution fit, at $\theta = 1 \ m/s^3$ and with $x_{min}$ estimated by power law fit.*

| dataset | distribution | normalized log-likelihood ratio | P value |
|---:|---|---:|---:|
| childplay | PLEC | 7.2 | 7E-13 |
| childplay | power law | 6.2 | 6E-10 |
| deskwork | PLEC | 2.1 | 0.03 |
| deskwork | power law | 2.1 | 0.03 |
| housework | PLEC | 2.7 | 0.01 |
| housework | power law | 2.5 | 0.01 |
| rest | PLEC | 3.9 | 1E-4 |
| rest | power law | 1.8 | 0.1 |

The likelihood comparisons produce even higher ratios when all of the data points are considered in the fitting, i.e. $x_{min} = 0.3$, as seen in Table 4.

Table 4: Normalized Log-Likelihood Ratios of Power Law and PLEC fits to the corresponding exponential distribution fit, at $\theta = 1\ m/s^3$ and with fixed $x_{min} = 0.3$.

| dataset | distribution | normalized log-likelihood ratio | P value |
|---:|---|---:|---:|
| childplay | PLEC | 24.4 | 4E-131 |
| childplay | power law | 24.4 | 4E-131 |
| deskwork | PLEC | 10.5 | 1E-25 |
| deskwork | power law | 10.5 | 1E-25 |
| housework | PLEC | 17.9 | 6E-72 |
| housework | power law | 17.9 | 6E-72 |
| rest | PLEC | 30.8 | 2E-208 |
| rest | power law | 30.8 | 2E-208 |

The PLEC model fits appropriately predict a metric discrepancy between BP and BTI via their parametric results. We present the maximum-likelihood estimated PLEC parameters per dataset, and the corresponding BP and BTI values, in Table 5.

Table 5: PLEC parameter estimates for the four datasets at $\theta = 1\ m/s^3$ and their corresponding BP and BTI values.

| Dataset | $\hat{\alpha}$ | $\hat{\lambda}$ | $\hat{x}_{min}$ | BP | BTI |
|---:|---:|---:|---:|---:|---:|
| childplay | 1.79 | 0.06 | 0.93 | 0.12 | 0.39 |
| deskwork | 3.24 | 0.01 | 1.33 | -0.09 | 0.37 |
| housework | 2.21 | 0.08 | 3.71 | -0.24 | 0.20 |
| rest | 1.11 | 0.53 | 0.93 | -0.24 | 0.09 |

# Appendix E

**Proposed Burstiness Quantitative Definition**

When researchers visually assess if an empirical distribution represents burstiness, they may plot the complementary empirical cdf against a comparable exponential distribution and qualitatively determined if the distribution "rolls off" more slowly than the comparable exponential distribution. This is not the same as the formal definition of heavy tails. Instead, we can formalize this "visual assessment" definition of burstiness as follows, for distributions with a finite mean:

Distribution F(x) is bursty $iff\ \exists \gamma\ such\ that\ \forall x > \gamma,\quad F(x) < \text{CDF}_{\exp}(x;\ \lambda_F, x_{min})$

where $\lambda_F$ is the MLE of $N$ samples from $F(x)$ as $N \to \infty$ (or equivalently, the $\lambda_F$ which minimizes the KL divergence of the empirical distribution and the exponential distribution) which is:

$$\lambda_F = \frac{1}{\mathbb{E}[X] - x_{min}}$$